\documentclass[12pt,preprint]{aastex}

\slugcomment{To appear in ApJ}
\usepackage{longtable}

\shorttitle{A Sparse Young Cluster, Lynds 988e}
\shortauthors{Allen et al.}

\begin{document}

\title{{\it Spitzer}, Near-Infrared, and Submillimeter Imaging \\of the Relatively Sparse Young 
Cluster, Lynds 988e}

\author{Thomas S. Allen\altaffilmark{1}, Judith L. Pipher\altaffilmark{1,2}, Robert A. 
Gutermuth\altaffilmark{3,2},  S. Thomas Megeath\altaffilmark{4,2}, Joseph D. 
Adams\altaffilmark{5}, Terry L. Herter\altaffilmark{5}, Jonathan P. Williams\altaffilmark{6}, 
Jennifer A. Goetz-Bixby\altaffilmark{1}, Lori E. Allen\altaffilmark{3}, Philip C. 
Myers\altaffilmark{3}}

\altaffiltext{1}{Department of Physics and Astronomy, University of Rochester, Rochester, NY 
14627}
\altaffiltext{2}{Visiting Astronomer, Kitt Peak National Observatory, National Optical Astronomy 
Observatory, which is operated by the Association of Universities for Research in Astronomy, Inc. 
(AURA) under cooperative agreement with the National Science Foundation.}
\altaffiltext{3}{Harvard-Smithsonian Center for Astrophysics, Mail Stop 42, 60 Garden Street, 
Cambridge, MA 02138}
\altaffiltext{4}{University of Toledo, Department of Physics and Astronomy, Toledo OH 43606}
\altaffiltext{5}{Cornell University, Department of Radiophysics Space Research, Ithaca NY 
14853}
\altaffiltext{6}{Institute for Astronomy, University of Hawaii, 2680 Woodlawn Drive, Honolulu, 
HI 96822}

\begin{abstract}

We present {\it Spitzer} images of the relatively sparse, low luminosity 
young cluster L988e, as well as complementary near-infrared (NIR) and submillimeter images of 
the region.  The cluster is asymmetric, with the western region of the cluster embedded within the molecular 
cloud, and the slightly less dense eastern region to the east of, and on the edge of, 
the molecular cloud. With these data, as well as with extant H$\alpha$ data of stars primarily found in the eastern region of the cluster, and a molecular $^{13}$CO gas emission 
map of the entire region, we investigate the distribution of forming young stars with respect to the cloud material, concentrating particularly on the differences and similarities between the exposed and embedded regions of the cluster. We also
compare star formation in this region to that in denser, more luminous and more massive clusters 
already investigated in our comprehensive multi-wavelength study of young clusters within 1 kpc 
of the Sun.

\end{abstract}


\section{INTRODUCTION}

Lynds 988e (L988e =IRAS 21023+5002) was identified as one of six IRAS point sources (L988a-f) 
observed toward the L988 dark cloud, which is located on the edge of the Cygnus OB7 molecular 
cloud association \citep{dobashi94} and which contains several bright, pre-main sequence objects. 
Bipolar molecular outflows were discovered \citep{clark86} to be associated with the vicinity of 
L988e, as well as the vicinity of L988a and L988f, and there is also high velocity CO emission (blue-shifted 
lobe only) to the west of L988c. There are two LkH$\alpha$ objects and related nebulosity seen 
both at visible and infrared wavelengths, LkH$\alpha$ 324-SE and LkH$\alpha$ 324, the brightest 
cluster members of a region of $\sim$50 H$\alpha$ emission-line stars \citep{herbig06} on the 
edge of the heavy extinction associated with  L988e's parent molecular cloud.  Most of these 
emission-line stars lie to the east of the molecular cloud or on its eastern edge.   LkH$\alpha$ 324 is 
a Herbig Ae/Be (HAeBe) star, close (in projection) to the IRAS coordinate, and LkH$\alpha$ 324-SE (misidentified as LkH$\alpha$ 324 in the the Herbig and Bell catalog \citep{herbig88}) is probably also an HAeBe star \citep{herbig06}. Both Classical T Tauri stars (CTTS) and Weak-line 
T Tauri stars (WTTS) are included in the emission-line stellar population.  Employing an average 
extinction of A$_V \sim$2.5 mag for most of the population, and measured extinctions for a few 
objects, \citet{herbig06} estimate median ages of 0.6 and 1.7 Myr for the emission-line stellar 
population from $V$ and $I$ photometry: for these estimates they assume evolutionary isochrones 
from \citet{DM97}, and \citet{siess00} respectively.  In this paper, we examine properties of the
entire cluster, including both the emission-line stellar population and an obscured population to the west within the molecular cloud which we have
identified here, and refer to as the embedded population.

At a distance of $\sim$700~pc and a far infrared luminosity of $\sim200~L_{\odot}$, the cluster of 
young pre-main sequence stars associated with L988e is relatively sparse, and in the bottom 
quartile in far-infrared luminosity of the northern young cluster sample surveyed in $^{13}$CO 
and C$^{18}$O molecular gas \citep{ridge03} (see Table~\ref{littable}). Despite present interest 
in young cluster evolution as discussed, for example, in a recent review \citep{ll03}, with the 
exception of papers by  \citet{hodapp94} and  \citet{herbig06}, the young cluster associated with 
L988e has been relatively unstudied at infrared wavelengths to date. Hereafter we refer to the 
young cluster as the L988e cluster, or just L988e. 

\begin{figure}
\epsscale{.9}
\plotone{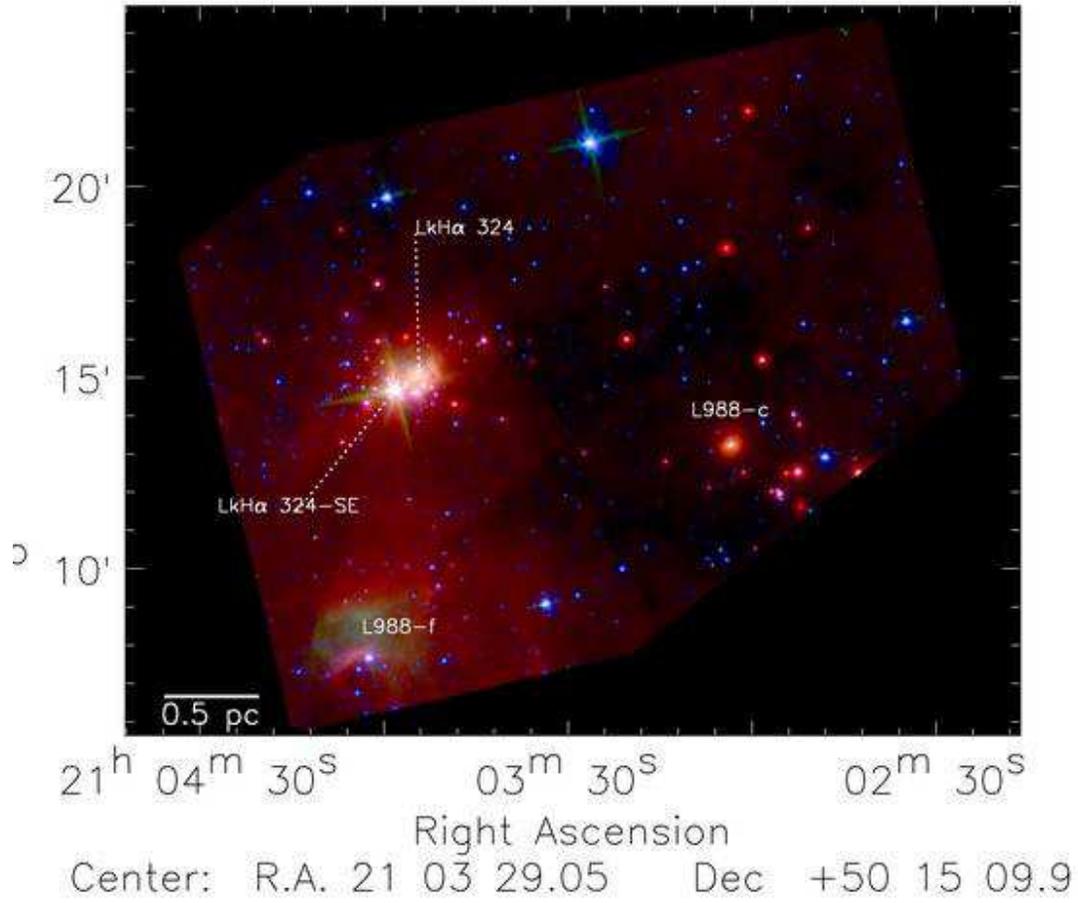}
\caption{Three-color 3.6, 5.8, and 24 $\mu$m image of the young asymmetric cluster L988e, including regions 
f and c.  The HAeBe stars LkH$\alpha$ 324 and 324SE are found in the L988e region.  
\label{13mips_fig}} 
\end{figure}

\begin{figure}
\epsscale{.9}
\plotone{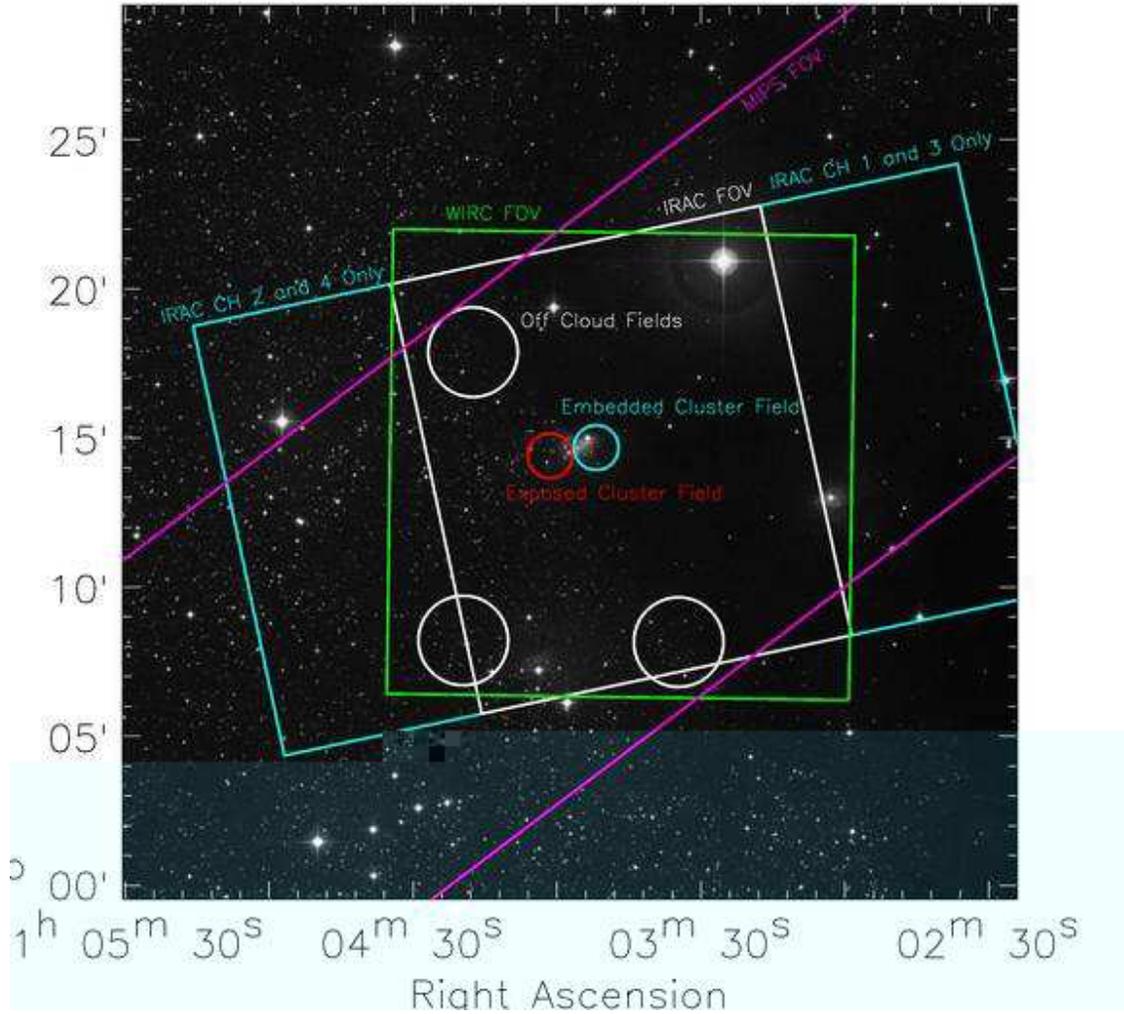}
\caption{Palomar Observatory Sky Survey II V-band image with the rectangular fields of view for 
the data presented in this work superimposed.  Additionally, the embedded (blue) and exposed (red) 
regions of the cluster and the off-cluster fields (white) used when determining disk fractions are shown in 
this figure as circles.  \label{cover_fig}} 
\end{figure}

In this paper, we utilize mid-infrared (MIR) images of L988e obtained with the Infrared Array 
Camera (IRAC) and 24 $\mu$m images obtained with the Multi-Band Imaging Photometer (MIPS) 
on the {\it Spitzer} Space Telescope\footnote{This work is based [in part] on observations made with 
the Spitzer Space Telescope, which is operated by the Jet Propulsion Laboratory, California 
Institute of Technology under a contract with NASA.} (see a composite three-color {\it Spitzer} image of the L988e region, also including L988f and c in Fig.~\ref{13mips_fig}; also see Fig.~\ref{cover_fig} where the various fields of view are shown on a reproduction of the POSS V-band image of the region), as well as complementary near-infrared (NIR) ground-based images, to identify cluster members via their infrared excess emission 
from protostellar envelopes, and from circumstellar disks.  The {\it spatial 
distribution} of identified protostars and disk objects in L988e is compared with the 850 $\mu$m 
continuum maps of cold dust emission obtained with Submillimeter Common-User Bolometer 
Array (SCUBA) on the James Clerk Maxwell Telescope (JCMT), with dust extinction maps 
generated using NIR data, and with a $^{13}$CO map of the associated molecular clouds 
\citep{ridge03}. These data were gathered as part of a coordinated multi-wavelength Young Stellar 
Cluster Survey, with the goal of providing the spatial distribution of a complete census of YSOs -- 
young stellar objects (all objects with an infrared excess, including stars with disks and protostars) 
in $\sim$30 nearby embedded clusters chosen from a list of 73 young groups and clusters within 1 
kpc of the Sun \citep{porras03}. Program clusters are representative of quite diverse star-forming 
regions which vary in number of cluster members, member number density, luminosity, and 
environment.  This coordinated study, which includes NIR, mid-infrared (MIR), submillimeter 
and millimeter CO imaging of the regions, is designed to advance our understanding of clustered 
star formation.  A comprehensive paper examining YSO spatial distributions in this survey is in 
preparation: in a few instances, where unique data and/or characteristics distinguish the object, a 
separate paper is written.  Here, the availability of deep NIR images and a published catalog of H$\alpha$ emission-line stars \citep{herbig06} distinguishes this object from the others in the survey.

Several authors have described efforts to define IRAC-only \citep{allen04}, or combined 
IRAC/MIPS \citep{muzerolle04} protocols which differentiate between stars with dusty 
circumstellar accretion disks (class II), and protostars or objects dominated by dust emission from 
envelopes (class I).  All class I and II objects exhibit infrared excess over photospheric emission 
typical of more evolved young cluster members (diskless or optically thin disk class III) as well as 
of foreground/background normal stars. Since substantial and/or variable interstellar 
extinction can cause ambiguity in identification of these objects using IRAC-only methods 
\citep{gutermuth05}, and the MIPS 24 $\mu$m images may not detect the faintest YSOs, a more robust classification method is required.  \citet{gutermuth05} used 
seven-band photometry from  $J$-band to IRAC 8.0 $\mu$m to systematically determine de-reddened SED slopes.  Using this classification as a standard, methods requiring five-band detection at $J$ (1.25 $\mu$m), $H$ (1.65 $\mu$m), $K_{s}$ (2.15 $\mu$m), [3.6] (3.6 $\mu$m), [4.5] (4.5 $\mu$m), or  four-band detection at $H$,$K_{s}$,[3.6], [4.5] also have been demonstrated in the same publication.  Here we primarily employ a combination of the five-band and four-band methods, but we also use slopes deduced from IRAC photometry in combination  with MIPS 24 $\mu$m photometry, where available, to classify protostars and transition disk 
objects (objects with a pronounced inner disk gap) in L988e. Alternate methods are exploited for the young cluster IC 348 \citep{lada06}: they utilize a simple power-law fit to the observed spectral energy distribution (SED) from the 4 IRAC bands to determine a slope 
$\alpha_{IRAC}$, and as well study the SED from visual to far-infrared wavelengths for all cluster members as a function of spectral class.  

\section{OBSERVATIONS \& DATA ANALYSIS\label{obs}}

\subsection{Near-Infrared Data\label{Obs_nir}}
NIR images in the $J$, $H$, and $K_{s}$ 
wavebands of L988e were first obtained by the authors on 2002 October 26 using the Simultaneous 
Quad Infrared Imaging Device (SQIID), which houses four $1024 \times 1024$ InSb arrays, of 
which only $512 \times 512$ are used, on the 2.1-m telescope at the Kitt Peak National 
Observatory, a facility of the National Optical Astronomical Observatory. The plate-scale of SQIID 
on the 2.1-m is $0^{\prime\prime}.68$ pixel$^{-1}$, and the circular field of view is 
$5^{\prime}.8$ in diameter.  SQIID images, obtained in an overlapping $3 \times 3$ grid 
pattern, were used to construct a mosaic comparable in size to the $15^{\prime} \times 
15^{\prime}$ {\it Spitzer} IRAC field of view.  Twelve dithered mosaics were obtained in each 
band with a total exposure time of 60~s per image at $J$, $H$, and $K_{s}$, leading to 720~s of exposure time.  With seeing, the typical point spread function (PSF) was $1^{\prime\prime}.9$ FWHM at $J$-band and $1^{\prime\prime}.77$ in the other two bands, and 
the 90\% integrated completeness limits of the final mosaics were estimated \citep{guter05} 
as $K_{s} = 16.19$, $H = 16.20$, and $J=17.71$.  However, the image quality suffered because all 
images exhibit position-dependent coma.  PSF photometry was consequently impossible, and 
aperture photometry with large apertures was impractical because of the crowded field.

Higher spatial resolution, deep, NIR observations in the $J$~(1.25~$\mu$m), $H$~(1.65~$\mu$m), and 
$K_{s}$~(2.15~$\mu$m) wavebands were obtained on 2004 July 6 with the Palomar Wide-field IR 
Camera (WIRC) which houses a $2048 \times 2048$ HgCdTe array \citep{wilson03}.  The plate-scale of WIRC is $0^{\prime\prime}.249$ pixel$^{-1}$, and the field of view is $8^{\prime}.7$ on 
a side.  In addition, the PSF was $1^{\prime\prime}.1$ in the $K_{s}$ band and $1^{\prime\prime}.3$ 
in the other two bands. The observing sequence consisted of a large ($3^{\prime}.5$ offset) five-point dither pattern. The 90\% integrated photometric 
completeness limits of the final mosaics were estimated \citep{guter05} to be approximately 
$K_{s} = 17.21$, $H = 17.72$, and $J =18.22$.  

Basic reduction of the SQIID NIR data was performed at Rochester using 
custom IDL routines, developed by the authors, which include modules for 
linearization, flat-field creation and application, background frame creation 
and subtraction, distortion measurement and correction, and mosaicking.  Point 
source detection and synthetic aperture photometry of all point sources were 
carried out using PhotVis version 1.09, an IDL GUI-based photometry 
visualization tool \citep{gutermuth04}.  By visual inspection, those detections 
that were  identified as structured nebulosity were considered non-stellar and
rejected. Aperture photometry was performed using radii of 
$2^{\prime\prime}.04$, $3^{\prime\prime}.40$, and $5^{\prime\prime}.44$ for the 
aperture, inner, and outer sky annular limits, respectively.  

WIRC data reduction was completed at Cornell using custom IDL routines and in-house image 
processing software. Processing steps included dark subtraction, linearization, sky subtraction, flat-fielding, spatial flux gradient correction (due to a dependency of the photometric zero point on 
position on the array; Jarrett personal comm.), and mosaicking. Aperture photometry using APPHOT 
was applied to all point sources.  Radii of the apertures and inner and outer limits of the sky annuli were 
$1^{\prime\prime}.5$, $2^{\prime\prime}.75$, and $3^{\prime\prime}.5$ respectively.  PSF fitting was attempted, but the PSF varied sharply across the images, with resulting residuals larger than obtained for the aperture photometry discussed above. Aperture photometry with larger apertures was tested, but not applied because it failed in the crowded field.  WIRC photometry was calibrated by minimizing residuals to 
corresponding 2MASS detections for selected stars with magnitudes between $\sim$12-14 mag.  Outliers 
were rejected from the fit (outliers are confused or unresolved multiples in the 2MASS dataset).  
The rms scatter for the $J$, $H$, and $K_{s}$ waveband residuals are 0.06, 0.07, and 0.07 mag 
respectively.  \citet{carpenter01} showed that $\sim$7\% of YSOs toward Orion A are variable: 
variability may contribute to the observed scatter.  Photometric magnitudes for stars which were brighter than the range of 
magnitudes corrected by the linearization modules were replaced with 2MASS photometry.
 Because of the superior spatial resolution, depth, image 
quality, yet similar field of view of the WIRC dataset, we use it for classification for all sources in L988e, and include photometry in Tables ~\ref{classItable},~\ref{classIItable},~\ref{classtdtable}, and ~\ref{classIIItable}. 
The SQIID dataset was used only for internal verification, and is not tabulated here. For fainter magnitudes (14-16) after rejecting outliers beyond 3$\sigma$, the rms scatter of the $J$, $H$, and $K_{s}$ waveband residuals between the SQIID and WIRC datasets was 0.08, 0.11, and 0.11 mag respectively.   We believe the WIRC photometry to be more accurate for the reasons cited above.

\begin{figure}
\epsscale{.9}
\plotone{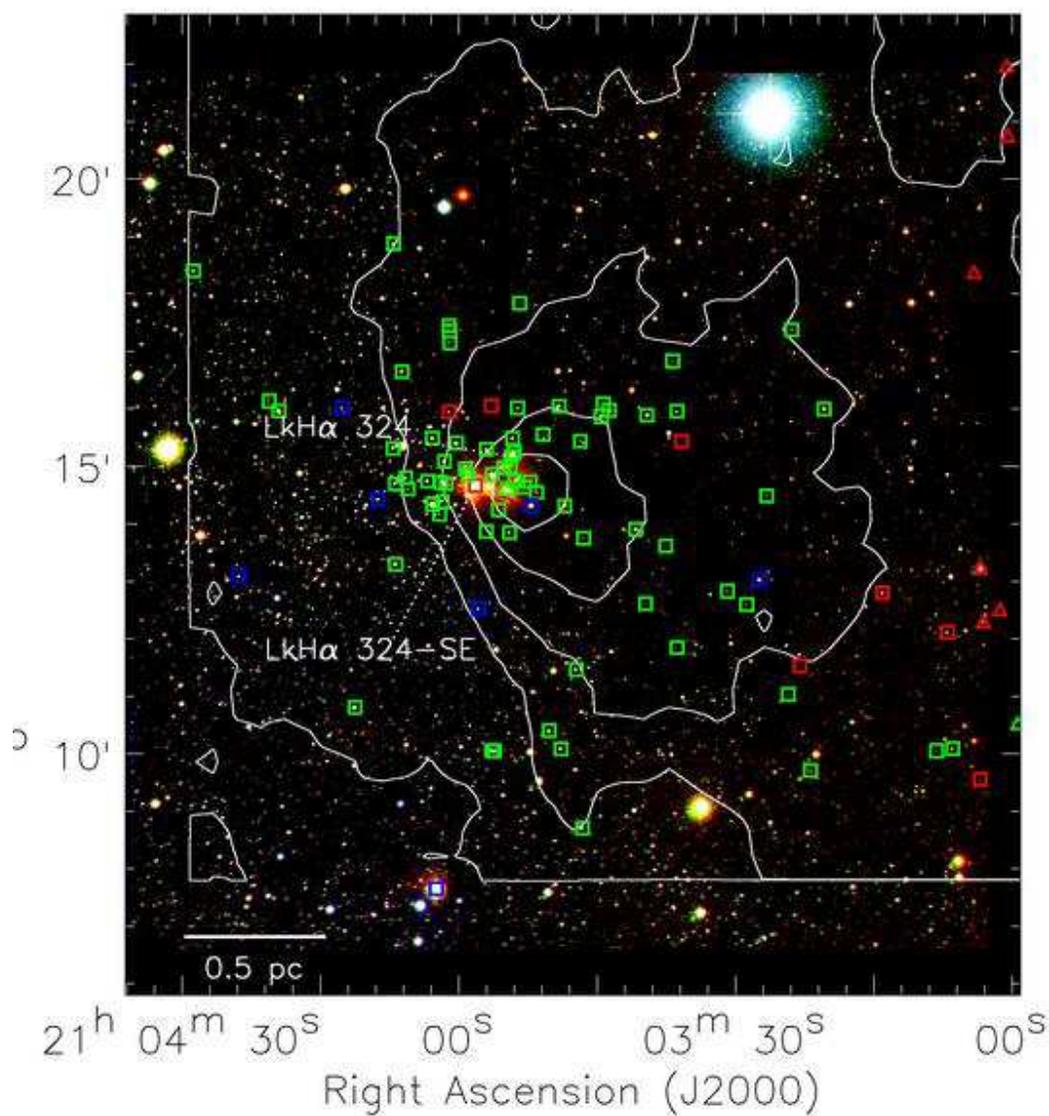}
\caption{Three-color $J, H, K_{s},$ image of the L988e region, with contours of $^{13}$CO  emission 
(50$^{\prime\prime}$ resolution) overplotted (\citep{ridge03}).  The $^{13}$CO  emission has a 
maximum integrated intensity of 13.75 K km s$^{-1}$ and contours range from 10\% to 90\% of 
the maximum in increments of 20\%. Squares designate objects classified via the 
($J$)$H$,$Ks$,[3.6],[4.5] methods or IRAC+MIPS methods and triangles designate objects 
classified via IRAC  [3.6] and [4.5] and MIPS [24] photometry.  Red designates identified 
protostars, green designates identified stars with disks, and blue designates identified stars with 
transition disks.  Note the paucity of stars detected toward the molecular cloud in the image. 
\label{nir13CO_fig}} 
\end{figure}

Fig.~\ref{nir13CO_fig} shows a three-color (WIRC $J$, $H$, $K_{s}$) image of the region with 
$^{13}$CO line emission contours \citep{ridge03} overplotted.  Visual inspection of the figure shows that 
the stellar density to the east of the molecular cloud is considerably higher than within the extent of 
the molecular cloud, as expected. 

\subsection{Mid-Infrared Data\label{Obs_mir}}
Mid-infrared images were obtained with {\it Spitzer}/IRAC at 3.6, 4.5, 5.8, and 8.0 $\mu$m and 
{\it Spitzer}/MIPS at 24 $\mu$m as part of the {\it 
Spitzer} Young Stellar Cluster Survey.  The IRAC images were obtained 2004 July 27, and each 
image was $5^{\prime}$ on a side with $1^{\prime\prime}.2$ 
pixels.  They were taken in 12 s High Dynamic Range (HDR) mode, whereby a 0.4 s 
integration is taken, immediately followed by a 10.4 s 
integration at the same pointing.   Four dithered  3 $\times$ 4 mosaics were obtained at each 
wavelength.  The typical PSF is approximately $2^{\prime\prime}.2$ at 3.6 and 4.5 $\mu$m, 
$2^{\prime\prime}.3$ at 5.8 $\mu$m, and $2^{\prime\prime}.5$ at 8.0 $\mu$m.  Ninety percent 
completeness limits of the final mosaics are approximately $[3.6] = 14.68$, $[4.5] = 14.67$, $[5.8] = 13.72$, and $[8.0] = 12.73$, and were evaluated using the same estimation technique as used for the NIR data.  
The {\it Spitzer}/MIPS 24 $\mu$m images were obtained on 2003 December 29, using scan mode 
at a medium scan rate.  The total area imaged is about $20^{\prime} \times 1^{\rm o}$.  The 
typical point spread function is approximately $6^{\prime\prime}.1$, and the 90\% completeness 
limit is $[24] = 8.18$. 

Basic reduction of the {\it Spitzer}/IRAC data was executed using the {\it
Spitzer} Science Center's
Basic Calibrated Data (BCD) pipelines and {\it Spitzer}/MIPS data using
the Post-BCD
product.  The IRAC and MIPS Data handbooks\footnote{The IRAC and MIPS Data
Handbooks can be found at
http://ssc.spitzer.caltech.edu/irac/dh/ and
http://ssc.spitzer.caltech.edu/mips/dh/, respectively.} give complete
descriptions of the calibration processes for point sources.  The IRAC
Data Handbook performs calibration photometry on a set of isolated standard stars
using a 10 pixel ($12^{\prime\prime}$) aperture, and 10 and 20 pixel radii
of the sky annulus, but notes that considerably smaller apertures are
required in crowded fields, and that small off-source sky annuli are
needed when the point sources are embedded in extended emission.  In those
cases, aperture corrections are required, and the Data Handbook 3.0
provides aperture corrections for various standard aperture and annuli
configurations for pipeline version S13. The same considerations apply to
the MIPS data.

Post-pipeline processing of IRAC data was performed at Rochester using
custom IDL routines,
developed by the authors, which include modules for bright source artifact
correction, cosmic ray
removal and mosaicking.  Point source extraction and aperture photometry
were performed using
PhotVis version 1.09.  Radii of $2^{\prime\prime}.4$,
$2^{\prime\prime}.4$, and
$7^{\prime\prime}.2$ were used for the aperture, inner, and outer sky
annular limits, respectively,
for the IRAC data and radii of $7^{\prime\prime}.4$,
$14^{\prime\prime}.7$, and
$24^{\prime\prime}.5$ were used for the aperture, inner and outer sky
annular limits, respectively,
for the MIPS data.  For a Gaussian PSF, an aperture radius equal to the
FWHM contains 93.7\% of
the total flux.   Increasing this aperture would negligibly gain signal
from the source and increases
total noise from the substantial MIR background emission.  The two
short-wave IRAC detector channels, 3.6 and 4.5 $\mu$m respectively, have
approximately Gaussian PSFs (other than the effects of diffraction), so
that choosing an aperture radius approximately equal to the observed FWHM
implies only a small correction because of the small aperture.  Point
source images for the two long-wave IRAC detector channels, 5.8 and
8.0 $\mu$m respectively, do not have a Gaussian PSF: much of the point
source intensity is
scattered in bands along rows and columns, as well as over the entire
array and beyond (see Data Handbook and \citet{pipher04}), but is
well-characterized (see  \citet{reach05}).  The chosen aperture of
$2^{\prime\prime}.4$ only gives 63\% of the source flux from a 12
$\arcsec$ aperture in the 8.0 $\mu$m band.  In this band, PAH emission
nebulosity is very bright and often structured, necessitating use of
relatively small apertures and sky annuli.
IRAC
photometry was calibrated using zero-magnitude flux densities of $280.9
\pm 4.1, 179.7 \pm 2.6,
115.0 \pm 1.7$, and $64.13 \pm 0.94$ Jy in the 3.6, 4.5, 5.8, and 8.0
$\mu$m bands respectively
\citep{reach05} in a 10 pixel ($12^{\prime\prime}$) aperture, and aperture
correction factors of 1.213, 1.234, 1.578, and 1.584 respectively as
tabulated in the Data Handbook.  MIPS photometry was calibrated using a zero-point flux of 7.3 Jy, referenced to Vega 
\citep{mips_man}.

\subsection{Submillimeter Data\label{Obs_submm}}
L988e was observed with the SCUBA submillimeter camera at 850 $\mu$m on the James Clerk 
Maxwell Telescope\footnote{The James Clerk Maxwell Telescope is operated by The Joint 
Astronomy Centre on behalf of the Particle Physics and Astronomy Research Council of the United 
Kingdom, the Netherlands Organisation for Scientific Research, and the National Research Council 
of Canada.} (JCMT) on Mauna Kea, Hawaii over the course of 5 nights from March to August 
2003 in relatively poor weather conditions ($\tau$(225 GHz) = 0.25-0.5). The data were obtained in 
scan-mapping mode whereby a map is built up by sweeping the telescope across the source while 
chopping the secondary at a series of throws, $30^{\prime\prime}$, $44^{\prime\prime}$, and 
$68^{\prime\prime}$. To minimize striping artifacts, each set of three sweeps was performed 
alternately in right ascension and declination. The SURF reduction package \citep{jlh98} was used 
to flat-field the data, remove bad pixels, and make images. Although effective at removing the high 
sky and instrument background, this observing technique necessarily loses all features on scales 
larger than the largest chopper throw, $68^{\prime\prime}$. The calibration was performed by 
using the same observing technique to observe one or more of Mars, Uranus, and standard source 
CRL 2688. Fig.~\ref{scuba13CO_fig} presents the SCUBA map in gray-scale, with $^{13}$CO 
contours and YSOs (see below) overplotted.  This map is presented here for morphological 
comparison to cluster member spatial distributions.  

\subsection{Extinction Maps\label{Obs_ext}}
We derive the A$_{Ks}$ extinction maps using the $(H - K_{s})$ colors of {\it all stars} (11,313) detected in both 
bandpasses, using a method outlined in more detail by \citet{guter05}.  Briefly, the line of sight 
extinction to each point on a $2^{\prime\prime}$ grid throughout the region surrounding the L988e 
core was estimated using a variation of the nearest neighbor technique.  At each point the mean and 
standard deviation of the $(H$ - $K_{s})$ colors of the $N$ = 20 nearest stars to each grid point were 
calculated, using an iterative outlier rejection algorithm that would reject any star with an $(H - K_{s})$
color $>3\sigma$ from the mean, until the mean converged.  The assumed NIR portion of the reddening law
\citep{flaherty07} given in Table ~\ref{redtable} was used to convert the $(H - K_{s})$ value at each point to A$_{Ks}$, using A$_{Ks}$ = 1.82($(H - K_{s})$)$_{obs}$ - $(H - K_{s})$)$_{int}$), where $(H - K_{s})_{int}$ = 0.2 is assumed as an average intrinsic $(H - K_{s})$ color 
for stars in young clusters. Further, the $H$ - $K_{s}$ map was convolved with a $15^{\prime\prime}$ 
Gaussian kernel to match the SCUBA beam size.  We compare the resultant map of A$_{Ks}$ with 
the $^{13}$CO map from \citet{ridge03} in Fig.~\ref{ak13co_fig_ha}.  As can be seen, the WIRC-derived 
A$_{Ks}$ extinction map follows the general distribution of the molecular cloud as outlined by CO emission, 
but provides considerably greater spatial detail and extent.  

If stars within the molecular cloud are included in the above analysis,  a biased 
estimation of A$_{Ks}$ could result.  In regions where cluster members dominate the $N$=20 number counts, this 
estimation will be lower than the line of sight extinction through the entire cloud, and thsi deviation will be larger in regions of high extinction where stars beyond the cloud cannot be detected.  To account 
for this bias, the $(H - K_{s})$ derived extinction map has been augmented with an extinction map 
derived by estimating A$_{Ks}$ using SCUBA 850 $\mu$m flux, and utilizing a conversion factor of 
A$_{Ks}$/F$_{850}$ =0.272 mag pixel mJy$^{-1}$ and an assumed dust temperature of 30 K 
\citep{guter05}.  In areas where the SCUBA derived A$_{Ks}$ values exceed the $(H - K_{s})$ -
derived A$_{Ks}$ values by 30\%, the SCUBA derived values are used.   Only one small region 
coincident with the L988e molecular core required augmentation with SCUBA-derived extinction 
values in the WIRC extinction map.

\section{IDENTIFYING YOUNG STELLAR OBJECTS\label{identifying}}

\subsection{NIR and MIR YSO identification Methods\label{identify_01}}
\citet{gutermuth05} has shown that by de-reddening the NIR/IRAC colors of young cluster stars, 
the IRAC-only classification scheme \citep{allen04} can be modified to better isolate class II from 
class I objects.  Briefly, \citet{gutermuth05} reevaluated the original IRAC-only classification 
scheme by considering the SED slopes from the NIR to the MIR.  Slopes ($\alpha$ = d log$(\lambda 
F_{\lambda}$)/d log $(\lambda)$) were constructed for both measured photometry and de-
reddened photometry of point sources in the cluster field of view.  When de-reddened photometry 
was utilized on six clusters, histograms of numbers of stars 
vs. 2-8 $\mu$m SED slope separate into 3 bins corresponding to field stars or class III ($\alpha < -2$), class II ($-2 < \alpha  < -0.5$), and class I ($\alpha > -0.5 $), (see Fig.~\ref{yso_class}).  This figure is constructed using the reddening 
law discussed above \citep{flaherty07},  derived from IRAC and NIR cluster data: this reddening law proved measurably 
different from a previously published reddening law utilizing IRAC and NIR data \citep{indeb05} 
that is primarily associated with reddening from dust in the diffuse interstellar medium rather than 
from dust in a molecular cloud.  

In addition, \citet{gutermuth05} considered classification methods using $J$, $H$, $K_{s}$,[3.6],[4.5] photometry as well as $H$, $K_{s}$, 
[3.6], [4.5] photometry.  Where $J$-band detections are available, objects are de-reddened to the 
($J$ - $H$)$_o$ vs. ($H$ - $K_{s}$)$_o$ CTTS locus of \citet{mch97} in the former method, or lacking a $J$-band 
detection, to the ($H$ - $K_{s}$)$_o$ vs.([3.6] - [4.5])$_o$ CTTS locus from \citet{gutermuth05} in the 
latter.  \citet{gutermuth05} then applied the seven-band classifications to a de-reddened ($K$ - [3.6])$_o$ vs. ([3.6] - [4.5])$_o$ color-color diagram, where the class I and II objects proved to be well separated.  

Since the 5.8 and 8.0 $\mu$m photometry is less sensitive, we utilize de-reddened (see 
Table~\ref{redtable} for the assumed reddening law for young clusters) $(J)$, $H$, $K_{s}$, [3.6], [4.5] photometry to classify most objects via their 
location on the $(K - [3.6])_o$ vs. $([3.6] - [4.5])_o$ color-color diagram: these classifications are 
color-coded in Fig.~\ref{k12ccd_fig}.  Where 24 $\mu$m data are available, objects classified as 
class I that have $([5.8]-[24]) <$ 4 were reclassified as class II.  Additional class I and class II 
objects (hereafter class Is and class IIs, similarly class IIIs for class III objects) were identified utilizing SED slopes from IRAC and MIPS photometry, even when some objects were not detected in the $H$, $K_{s}$, or [3.6] bands. In addition, there are a handful of 
objects with NIR+IRAC colors of class IIIs, but which exhibit a 24 $\mu$m excess  (e.g. 
$([5.8] - [24]) >$ 1.5 and  $([8] - [24]) >$ 1.5): these objects are classified as transition disk objects, 
YSOs characterized by a cool circumstellar disk containing a substantial inner gap.  However, it 
must be noted that in the absence of a large 24 $\mu$m excess, it is difficult to distinguish 
primordial transition disk objects (with ages less than a few Myr) from regenerative debris disk 
objects (with ages of tens of Myr to a few Gyr).  Since the field of view covered by MIPS (as well 
as [3.6] and [5.8]) extends farther to the west than the $(J)$, $H$, $K_{s}$, [3.6], [4.5] field, additional 
YSOs are identified outside that field.

\subsection{Isolating Extragalactic Contamination\label{Identify_02}}
Deep NIR and MIR photometric detections of  YSOs  toward molecular clouds are contaminated by detection of extragalactic objects, including PAH (polycyclic aromatic hydro-carbon)-rich galaxies and active galactic nuclei (AGN), both of which have colors similar to YSOs.  We account for extragalactic contamination in the following way.  For objects which have detections in all four of the IRAC bands, we deredden each object on the assumption that it lies behind the molecular cloud, utilizing the extinction map discussed in $\S$2.4.  We then utilize ($[4.5] - [5.8]$) vs. ($[5.8] - [8.0]$) and ($[3.6] - [5.8]$) vs. ($[4.5] - [8.0]$) color-color diagrams to identify the PAH-rich galaxy contaminants and the $[4.5]$ vs. ($[4.5] - [8.0]$) color-magnitude diagram to identify the AGN contaminants.  The prescriptions for identification are described in detail by \citet{guter07}.  For objects which only have $(J)HK[3.6][4.5]$ photometry, we are unable to directly use those prescriptions to identify extragalactic contaminants.  Instead, inspection of the histogram of de-reddened $[3.6]$ magnitudes for the objects with an infrared excess utilizing the $(J)HK[3.6][4.5]$ classification scheme (Fig.~\ref{ch1_hist}) shows a bifurcation in the distribution at $[3.6]$ = 13.5.  Objects with  $[3.6]$  $<$ 13.5 are considered as potential YSOs, while objects with  $[3.6]$ $>$ 13.5 are considered as potential extragalactic contaminants and are rejected as potential YSOs.  This scheme is in agreement with previous work stating that at magnitudes fainter than about 14, extragalactic objects dominate number counts \citep{fazio04}.  In order to quantify this further, we examine the differential number counts for sources detected in the Bo\"{o}tes region in the Flamingos Extragalactic Survey \citep{elston06}.  At $[3.6]$ = 13.5 (Fig.~\ref{ch1_hist}), and a mean de-reddened $(K_s -[3.6])_o$ color of 0.79 (typical of YSOs), this bifurcation magnitude corresponds to $K_{s}$ =14.29.  From Fig. 8 in \citet{elston06},  we estimate that there are approximately 10 extragalactic sources within our field of view at $K_{s}$ =14.29, and the number rapidly increases to fainter magnitudes.  Since we are examining the spatial distribution as well as the disk fraction of the L988e population of YSOs, both in-cloud and off-cloud, we want to minimize the possibility of contamination by extragalactic objects.  An examination of the $K_{s}$ vs. $(H-K_{s})$ color-magnitude diagram (Fig.~\ref{kvkmh}) shows that even excluding $K_{s} \ge$14.29, we are still capable of detecting relatively low mass YSOs. Faint cluster members (below the hydrogen burning limit at A$_V$=0) will also be excluded by these criteria: therefore brown dwarf cluster members will not be considered further.  Spectroscopy is the best way to make firm brown dwarf identifications, and beyond the scope of this paper.

\subsection{Number Counts of Identified YSOs\label{Identify_03}}
Before removing extragalactic contaminants, there are a total of 2293 objects in a 14.7$^{\prime} 
\times 12.5^{\prime}$ field with detections in all four bands ($H$,$K_{s}$,[3.6],[4.5]) meeting the criterion that the photometric uncertainties are less than 0.1 mag in all 
bandpasses.  Following application of the extragalactic contamination criteria noted above, there remain a total of 678 objects in the NIR + IRAC dataset: these objects will be discussed in the following sections.  Of these, three are class Is, 69 class IIs, seven are 
transition disk objects, and 599 are class IIIs or field stars with no obvious 
infrared excess.  When SED slopes were computed for objects in the same field of view with only 
IRAC and/or MIPS detections, an additional five class Is, three class IIs, and 36 class IIIs were found.  Further, there is one object detected 
in only the 4.5, 5.8, 8.0, and 24 $\mu$m wavebands.  From the 4.5 - 24 $\mu$m SED slope for 
this object, $\alpha$=0.98, we characterize it as class I.  As discussed earlier, a total of six 
transition disk objects are also classified based on their ([5.8] - [24])  or ([8] - [24]) colors. In 
Fig.~\ref{IRAC_fig} we present a three-color IRAC image ($[3.6], [4.5], [8.0]$) of the L988e 
region, with WIRC + IRAC class I and II classifications overplotted, and in Tables ~\ref{classItable},~\ref{classIItable},~\ref{classtdtable}, and ~\ref{classIIItable} we present a catalog of identified objects.

The MIPS field of view extends far to the northwest of L988e, to include L988c, as noted above.  
Fortunately, IRAC data in two of the four bands (3.6 and 5.8 $\mu$m) were obtained to the 
northwest of L988e.  From 3-24 $\mu$m SED slopes in combination with NIR data, a total of 19
 class Is, 73 class IIs, and 643 class IIIs/field stars were found in the 
14.7$^{\prime} \times 19.4^{\prime}$ field.  A three-color 
IRAC + MIPS image ($[3.6], [5.8], [24]$) of the extended L988e region was presented in Fig.~\ref{13mips_fig}.  

\section{THE H$\alpha$ EMISSION LINE PRE-MAIN SEQUENCE 
POPULATION\label{Halpha}}

Early studies of the emission-line stellar populations in molecular clouds led to the classical T 
Tauri star (CTTS) and the weak T Tauri star (WTTS) definitions \citep{herbig88} in terms of their 
H$\alpha$ equivalent widths EW[H$\alpha$]. Although the WTTSs usually show weaker 
H$\alpha$ emission, limited both in level and profile width (from chromospheric emission), CTTSs 
exhibit strong emission lines, with broad profiles, generated in the accretion process.  By 1988, the 
CTTS classification was taken to be EW[H$\alpha$] $\ge$ 10 \AA.  However, in a recent study of 
low mass stars, \citet{white03} suggest an empirical relationship for the critical EW[H$\alpha$] 
value as a function of spectral type for low mass stars. Using the presence of optical veiling to 
identify accretion, they propose the following improved scheme of CTTS classification for spectral classes later than K0: EW[H$\alpha$] $\ge$  3 \AA , for 
K0-K5; EW[H$\alpha$] $\ge$ 10 \AA , for K7-M2.5; EW[H$\alpha$] $\ge$ 20 \AA , for M3-M5.5; and EW[H$\alpha$] $\ge$ 40 \AA, for M6-M7.5.  In what follows, we first use the single equivalent width classification of CTTS/WTTS, and then examine whether this analysis would change appreciably with the \citet{white03} modification.

In Fig.~\ref{ak13co_fig_ha}, cluster stars, including the HAeBe emission-line stars in this cluster identified by \citet{herbig06}, are overplotted on an image of the extinction at $K_{s}$-band (see $\S$2.4).  Each object in Fig.~\ref{ak13co_fig_ha} is coded by YSO classification, deduced as outlined in $\S$3 above:   green squares (class II),  red squares (class I), blue squares (transition disk objects).  Filled squares denote YSOs that are H$\alpha$ emission-line objects. Of course, with H$\alpha$ emission line measurements, the degeneracy between field stars and pre-main sequence class III cluster members is broken, so that class IIIs with H$\alpha$ emission can be identified (blue $\times$'s).  The bright HAeBe stars LkH$\alpha$ 324-SE (V=15.36, A$_V \sim$ 10) and LkH$\alpha$ 324 
(V= 12.61, A$_V$ = 3.75) are identified by us as class I and class II, respectively.  Even though the $K_{s}$-band error for LkH$\alpha$ 324 is slightly above our error threshold (error = 0.108, threshold = 0.1) we chose to include this object because of its prominence in the cluster.  Most of 
the cluster associated with the emission-line stars lies in a region of low extinction to the 
east of the molecular cloud, or on the eastern edge of the cloud, where there is modest extinction, although there are bright, heavily 
extinguished members in the sample.  

In Fig.~\ref{slope_width_fig}, the H$\alpha$ equivalent width EW[H$\alpha$], a proxy for 
accretion strength \citep{herbig06}, is plotted against the de-reddened SED slope from IRAC 3.6 to 8.0 $\mu$m 
photometry for the 41 emission-line stars common to the list from \citet{herbig06} and to our NIR 
+ IRAC dataset with uncertainties $\le$ 0.1 mag in all four IRAC bands. A vertical line at EW[H$\alpha$] = 10 
{\AA} divides the plot into the single-value estimate of WTTS ($<$ 10 \AA) and CTTS ($\ge$ 10 
\AA). In their study of members of the young cluster IC 348, \citet{lada06} show that diskless stars (Class III) exhibit an IRAC determined SED slope of $<$ -2.56, stars with optically thick disks exhibit slopes $>$ -1.8, and they define sources with SED slopes between -1.8 and -2.56 to be  `anemic' disk objects (this class includes transition disk objects, optically thin disks etc.): we have adopted these parameters in Fig.~\ref{slope_width_fig}.  Generally, Class IIs (or Is) in L988e with 
optically thick disks show the strongest H$\alpha$ equivalent widths, as they do in IC 348 (see also \citet{hartmann98} 
who correlated $(K-L)$ excess emission from optically thick disks with large EW[H$\alpha$]).  In order to compare our results with earlier studies, we have indicated by color our Class I, II, and III classifications.  Although there is a general correlation of Class II/I with strong H$\alpha$ equivalent widths in L988e, the figure also shows three stars with H$\alpha$ equivalent widths $>$ 10 \AA, but slopes indicating class III according to the above definition, and three with H$\alpha$ equivalent widths clearly $<$ 10 \AA, but slopes indicating thick disks (Class II).  There are five objects in the anemic disk region. Since six objects do not obey the correlation described above, we now determine whether use of the single-valued H$\alpha$ equivalent width definition of CTTS/WTTS is the cause of the departure.   In the absence of spectroscopy for these 41 L988e cluster members, we made rough estimates of the emission line stars' spectral 
types from their measured ($V$-$I$) colors \citep{herbig06} for objects where good $V$ and $I$ photometry was available. In order to do so, we first confirm that spectral types determined from $V$ and $I$ photometry can be roughly calibrated for stars with disks in young clusters.

The \citet{lada06} study of cluster members in IC 348 included published spectroscopic data, and they provide a table of median star-disk spectral energy distributions at wavelengths from  $V$-band to 24 $\mu$m, all with a reddening of A$_V \sim$ 2.5, as a function of nine spectral type intervals.  From their table, we confirm that the $(V - I)_o$ colors of both stars and objects with disks are sufficiently close for spectral typing into the nine intervals. We also examined the $JHK_{s}$ colors for the same median star-disk population in IC 348 \citep{lada06}, and found NIR colors to be unsatisfactory for rough spectral typing of YSOs, because, as expected, those colors can vary widely for photospheres, stars with `anemic disks' and stars with thick disks.  

Thirty-seven of the 41 emission line stars in L988e referred to above have good $V$ and $I$ photometry.  We first de-redden the $V$ and $I$ magnitudes using our determined values of A$_{Ks}$, and assuming A$_V$/A$_{Ks}$ = 10 (values of this ratio can vary by up to 30\% \citep{glass99}).  Although \citet{herbig06} used a single value for A$_V$ (2.5) with only a few exceptions, we found considerable variation in A$_{Ks}$ for these stars. Thus $(V - I)_o$ colors were estimated for these 37 emission-line stars, and 62\% are later than K7 as determined from those colors and the rough spectral type calibration.  

Applying the \citet{white03} criterion for CTTS/WTTS classification as a function of rough spectral type, we found that 6 of the 37 objects yielded a different CTTS/WTTS classification than with the single-valued H$\alpha$ equivalent width definition of CTTS/WTTS. These are indicated on Fig.~\ref{slope_width_fig} with the appropriate symbol encircled, and color-coded (green circle implies CTTS is the correct classification rather than WTTS, and a blue circle implies WTTS rather than CTTS). The resultant correlation of Class III with WTTS and Class II with CTTS is superior to that obtained using the single-valued H$\alpha$ equivalent width discriminator. Three transition disk objects are identified and are indicated by a blue square.  One of these is  an anemic disk and one is a thick disk object. The remaining transition disk object is found slightly below the -2.56 slope delineation, rather than above, so is marginally non-compliant with the correlation discussed above. There are two additional objects which fall in the anemic disk region. Overall, the correlation of CTTS/WTTS to Class II/Class III is excellent once the revision to the equivalent width CTTS/WTTS definition proposed by \citet{white03} is applied.

We now examine the statistics of the emission-line population.  In addition to the 41 objects 
discussed above, there are four H$\alpha$ objects which do not have detections in all four IRAC 
bands and six which do have detections, but with uncertainties in IRAC bands $>$ 0.1 mag.  There are also two emission-line objects in common with our IRAC (3.6 and 5.8 $\mu$m)  + MIPS 
dataset in the vicinity of L988c.  Of those objects with H$\alpha$ emission, 28 are class IIs, two are transition disk objects, 19 are class IIIs, and two are class Is.  Of the two in common with our IRAC (3.6 and 5.8 $\mu$m) + MIPS dataset, one is determined to be class I, and the other is determined to be class II from their 3 - 24 $\mu$m SED slopes and $([5.8]-[24])$ colors.

From these results, the disk fraction of {\it just} the emission-line population with NIR + IRAC 
counterparts is 67\% $\pm$20\% including five which could be described as `anemic disks' 
\citep{lada06} based on IRAC SED slopes or MIPS 24 $\mu$m detections.  This percentage is 
consistent with either of the two age estimates cited by \citet{herbig06}, namely 0.6 and 1.7 Myr, 
using the curve of disk fraction to age from \citet{hernandez07}.

\section{THE COOL DUST DISTRIBUTION\label{reddening}}

\begin{figure}
\epsscale{.9}
\plotone{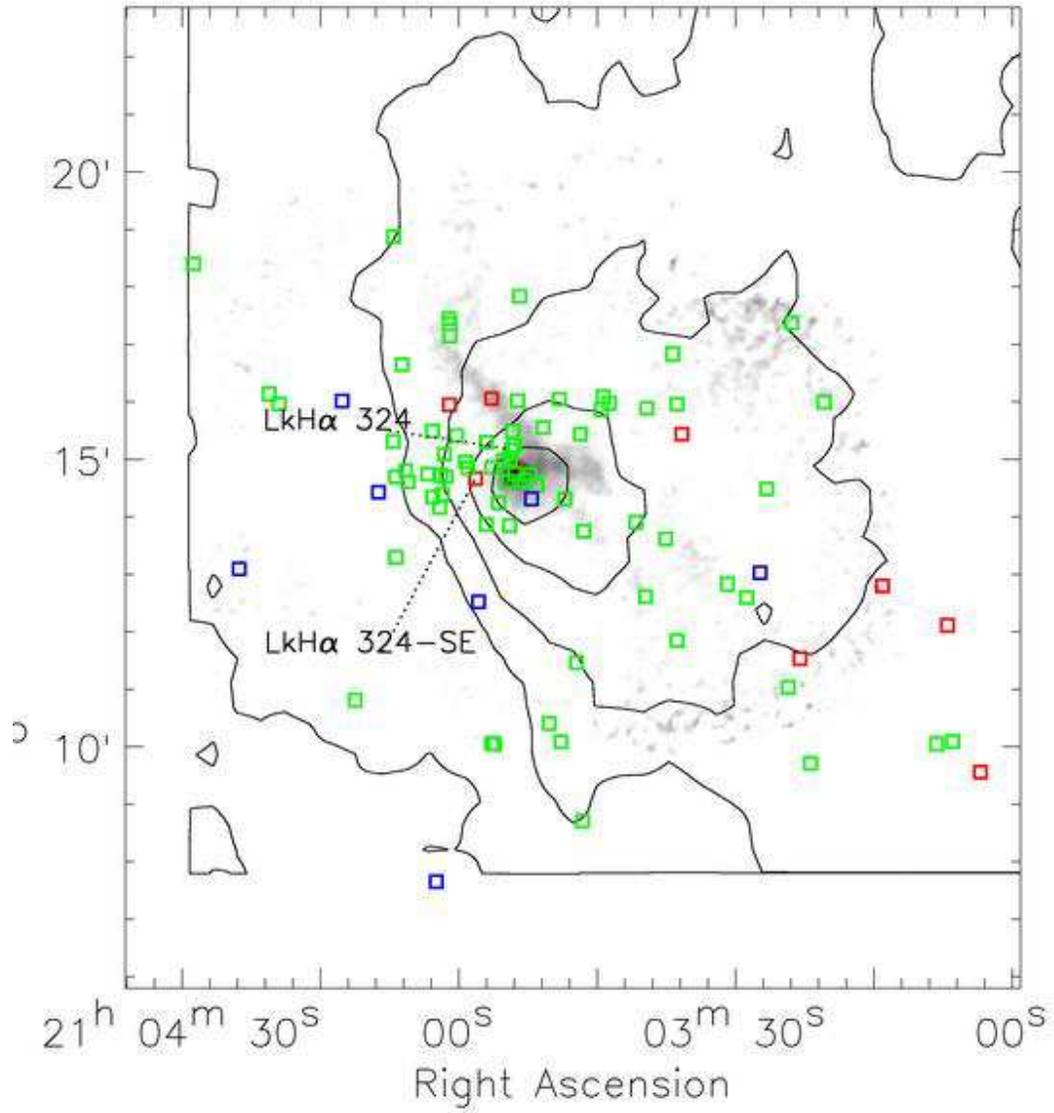}
\caption{Linear grey scale SCUBA 850 $\mu$m emission (15$^{\prime\prime}$ resolution), with 
contours of $^{13}$CO (50$^{\prime\prime}$ resolution) overplotted, with the same contours as 
in Fig. 3. Red designates identified protostars, green designates identified stars with disks, and blue 
designates identified stars with transition disks.\label{scuba13CO_fig}} 
\end{figure}

\begin{figure}\epsscale{.9}
\plotone{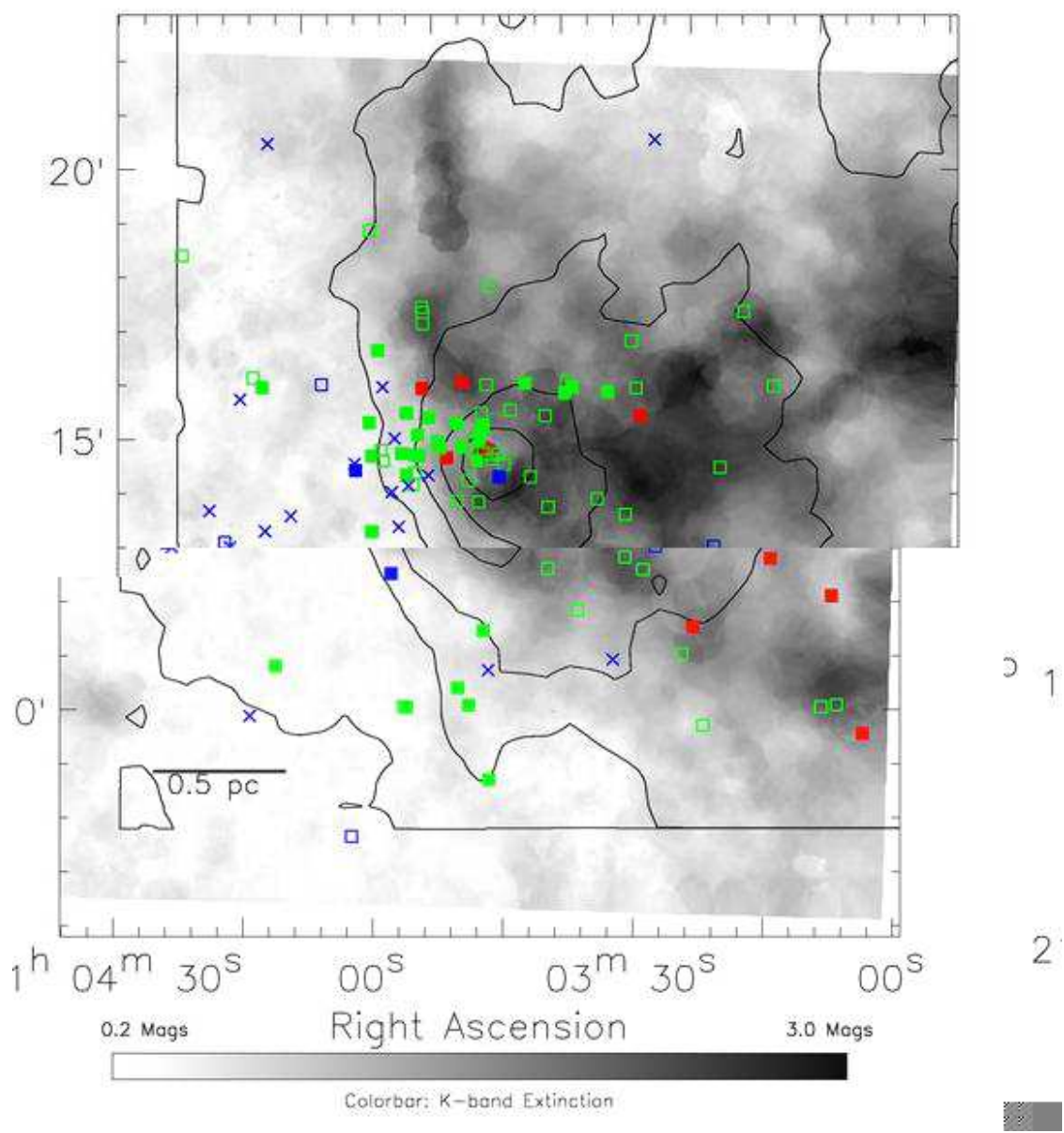}
\caption{A$_{Ks}$ extinction map from WIRC data in grey scale, with the same $^{13}$CO contours 
as in Fig. 3, and identified YSOs  overplotted.  The grey color scale extends from A$_{Ks}$ = 0.2 
(white) to 3 (black) mag.  In the region of the maximum CO intensity, SCUBA derived values are 
substituted for A$_{Ks}$.  Identified YSOs are plotted with red squares denoting class Is, green 
squares denoting class IIs, blue squares denoting transition disk objects, and blue $\times$'s denoting class 
IIIs with H$\alpha$ emission.  The YSO identified by filled symbols are H$\alpha$ emission-line 
stars \citep{herbig06}.  \label{ak13co_fig_ha}} 
\end{figure}

\begin{figure}
\epsscale{1.}
\plotone{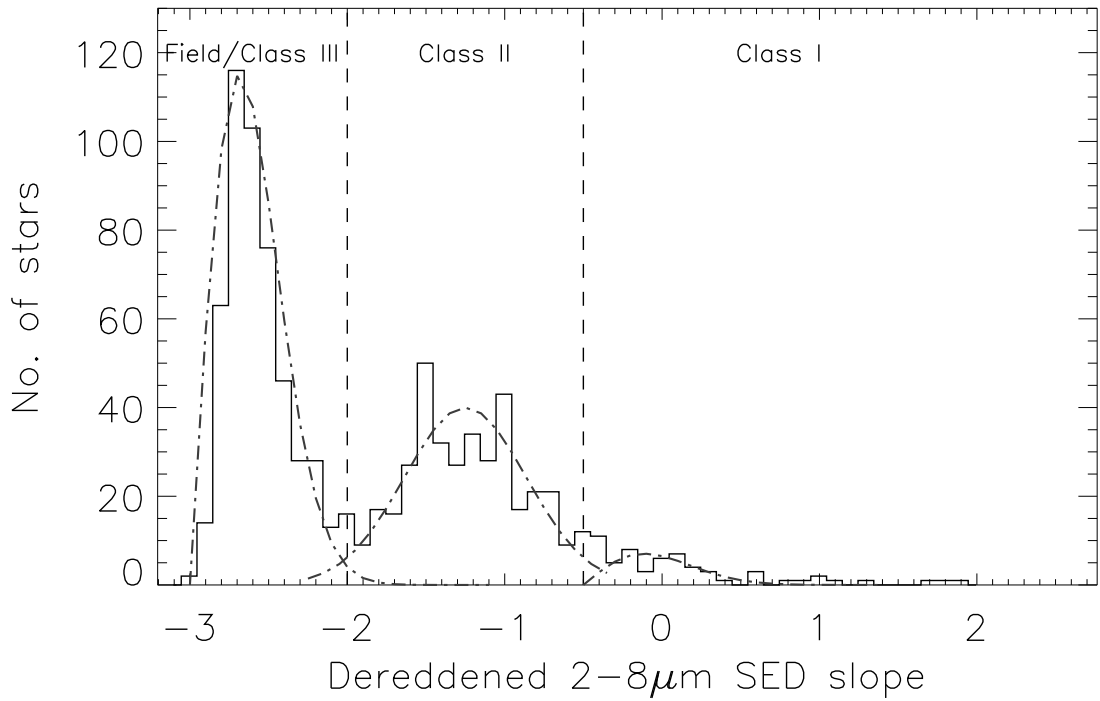}
\caption{Histogram of the number of objects vs. de-reddened 2 - 8 $\mu$m slope for six young 
clusters included in \citet{gutermuth05}.   Class III/Field stars are well separated from class IIs, 
and class Is are defined by consideration of distribution shapes. \label{yso_class}} 
\end{figure}

\begin{figure}
\epsscale{1.}
\plotone{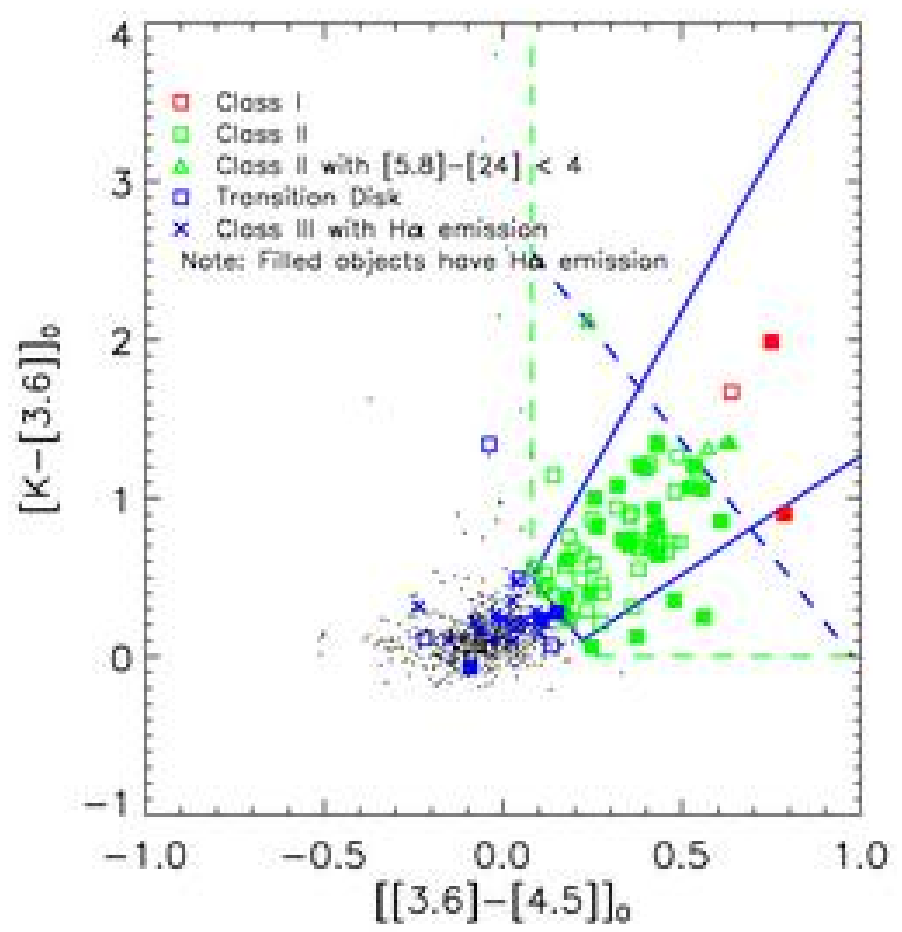}
\caption{Dereddened $(K_{s} - [3.6])_o$ vs. $([3.6] - [4.5])_o$ color-color diagram. Black dots represent either class 
III or field stars.  A dashed blue line shows the separation between class I and II. Solid blue lines 
show the limits of regions clearly dominated by class I and II objects, as given by de-reddened 2-8 
$\mu$m SED classifications, while the dashed green lines show the less restrictive region actually 
classified here, allowing for YSO variability.  Triangles denote objects originally classified as class 
I based on dereddened $(K_{s} -[3.6])_o$ vs. $([3.6]-[4.5])_o$ colors, but were reclassified as class II due to a $([5.8]-
[24])$ color of less than 4.\label{k12ccd_fig}} 
\end{figure}

\begin{figure}
\epsscale{.8}
\plotone{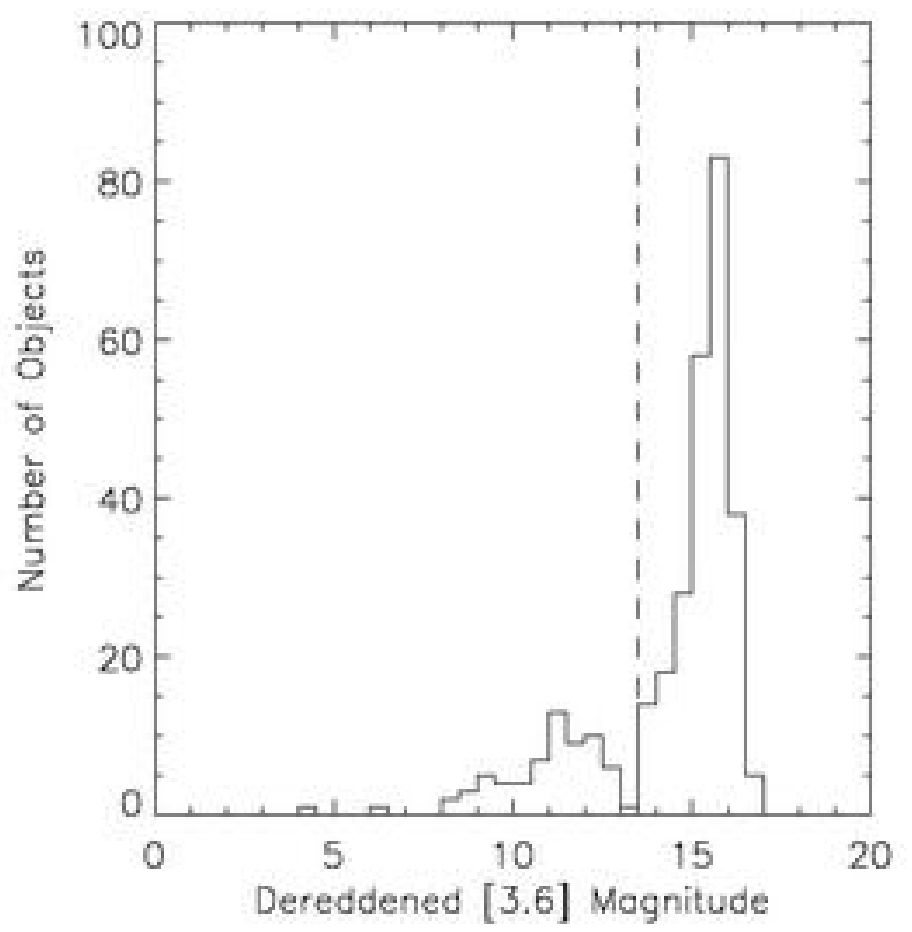}
\caption{Histogram of the number of stars vs. de-reddened 3.6 $\mu$m magnitudes for all objects 
classified as having an infrared excess from ($J$)$H$,$K_{s}$,[3.6], [4.5] photometry.  Note the 
separation between YSOs ([3.6] $<$ 13.5) and extragalactic contaminates ([3.6] $>$ 13.5). 
\label{ch1_hist}} 
\end{figure}

\begin{figure}
\epsscale{.8}
\plotone{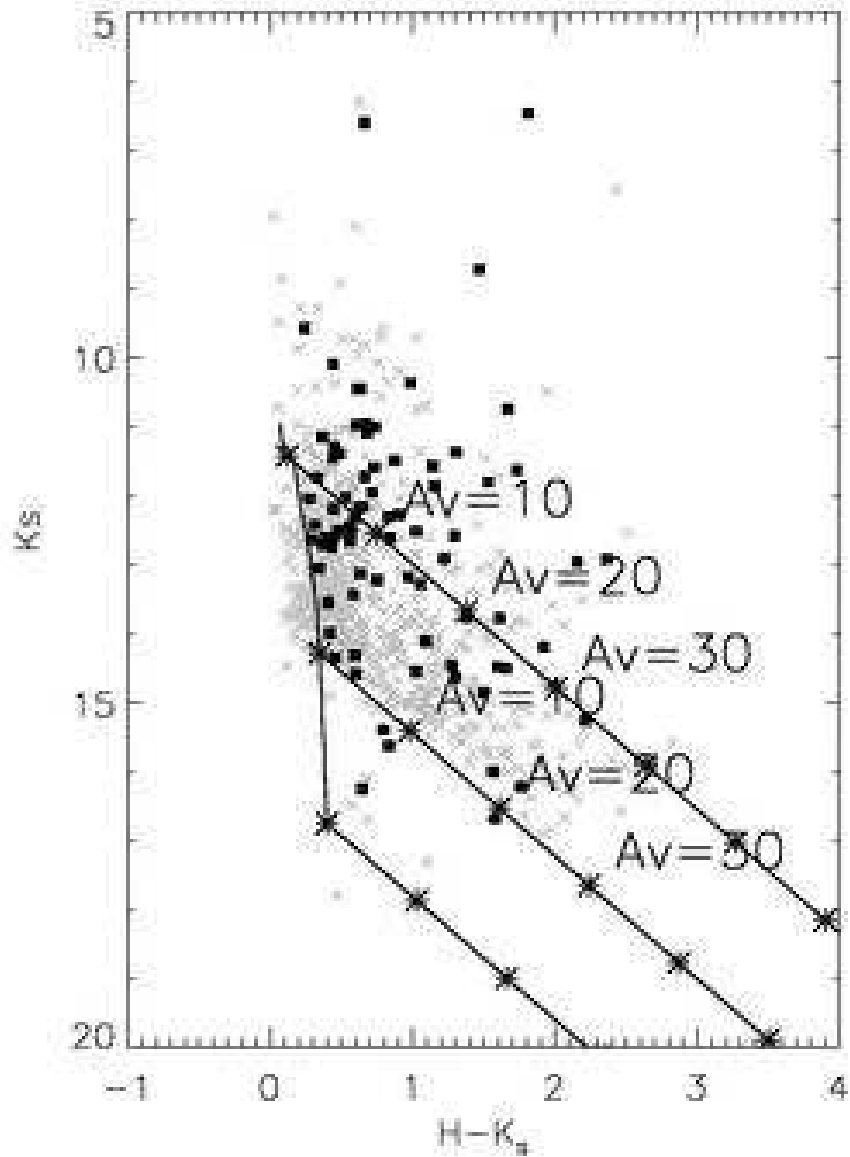}
\caption{NIR Color-Magnitude diagram for cluster members (extragalactic contaminants removed) of the L988e region with the 1 Myr isochrone of \citet{baraffe98}     overlaid in black.  Solid lines are the reddening vectors for a $1~M_{\odot}$ object (top), $0.08~M_{\odot}$ object, the Hydrogen Burning Limit (middle), and $0.025~M_{\odot}$ object (bottom).  Gray $\times$'s represent objects classified as class III/photospheric and black squares represent objects classified as YSOs by a $J,H,K_s,[3.6],[4.5]$ classification. \label{kvkmh}} 
\end{figure}

\begin{figure}
\epsscale{.9}
\plotone{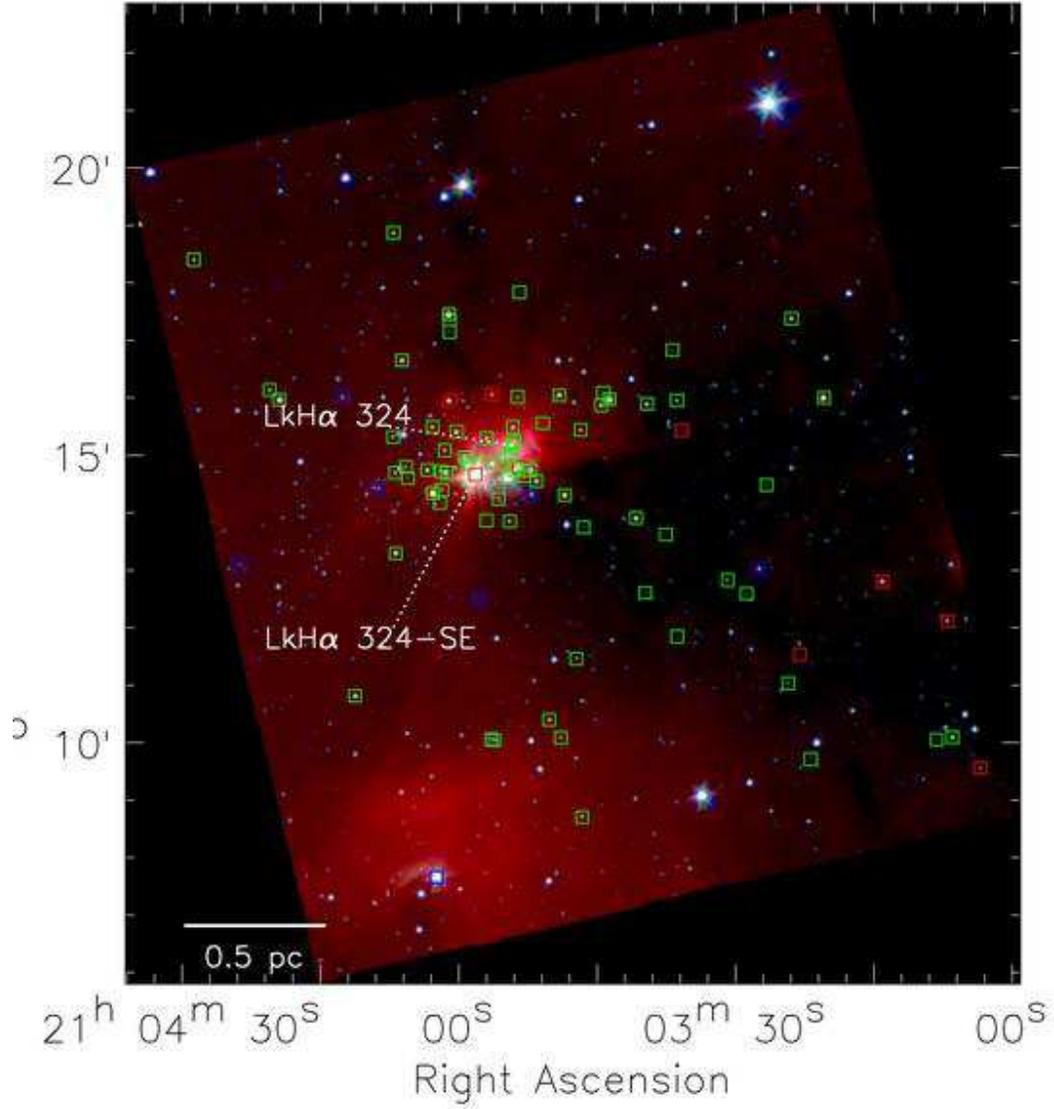}
\caption{Three-color [3.6], [4.5], [8.0] image of the L988e star formation region.  
Overplotted on the image are stars with disks (green boxes), transition disks (blue boxes), and 
protostars (red boxes). \label{IRAC_fig}} 
\end{figure}

\begin{figure}
\epsscale{.8}
\plotone{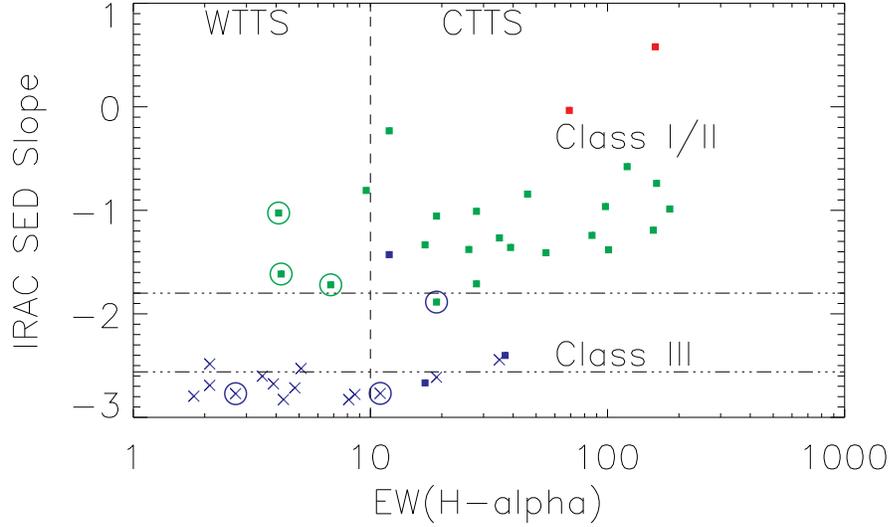}
\caption{De-reddened SED slope (from 3.6 $\mu$m to 8 $\mu$m) vs. EW[H$\alpha$] in 
\AA.  Here red squares represent class Is, green squares class IIs, blue squares transition disks and 
blue $\times$'s class IIIs determined from ($J$),$H$,$K_s$,$[3.6],[4.5]$ photometry.  A vertical dashed line 
at 10 \AA\ illustrates the accepted single-valued equivalent width division between WTTS ($<$ 10 \AA) and CTTS ($>$ 10 \AA).  Circled objects are those that would yield a different CTTS/WTTS classification using the \citet{white03} criterion for CTTS/WTTS classification as a function of spectral type.  Green circles imply CTTS rather than WTTS as determined from the single-valued equivalent width division, and blue circles imply WTTS rather than CTTS.  Two horizontal dot-dashed lines at slopes of -1.8 and -2.56 delimit the `anemic' disk region. Sources with thick disks fall above the -1.8 line and sources without disks fall below the -2.56 line. As demonstrated by \citet{lada06}, sources with disks (and envelopes) have stronger H$\alpha$ equivalent widths. \label{slope_width_fig}} 
\end{figure}

Young cluster environments are permeated by dust associated with the natal molecular cloud, 
providing extinction as well as the cool dust emission detected at submillimeter wavelengths.  All of the figures describing the molecular cloud will be used in the following discussion of the spatial distribution of the identified YSOs (Figs.~\ref{nir13CO_fig}, \ref{scuba13CO_fig}, and \ref{ak13co_fig_ha}).  The 
dust and molecular gas are distributed non-uniformly, and in order to characterize that distribution 
for comparison with the YSO distribution, we map the extinction (A$_{Ks}$) in the neighborhood of L988e.  
The peak of the SCUBA 850~$\mu$m emission (Fig.~\ref{scuba13CO_fig}) coincides with the 
peak intensity of $^{13}$CO emission, although a NE extension of the submillimeter emission 
does not have a parallel in the low spatial resolution molecular map: since the intensity of the 
emission varies on scales larger than the map size (~$5^{\prime}$) outside the peak, there is no 
information on the larger scale dust emission due to the iterative way the map is built, assuming 
that the edges chop into zero emission.  Thus the SCUBA-measured emission is not ideal for 
illustrating the extent of the molecular cloud.  Despite adequately mapping the extent of the cloud, 
available low spatial resolution ($50^{\prime\prime}$ beam size) CO emission line maps 
\citep{ridge03} are also not ideal for comparing the dust distribution with the YSO distribution.  In 
addition to inadequate spatial resolution, there are issues associated with low signal to noise in 
diffuse regions (C$^{18}$O), and optical depth issues leading to saturation in dense cores 
($^{13}$CO), as well as CO depletion due to chemistry and/or freeze-out in dense cores, and/or 
varying excitation temperature. 

Although we could in principle utilize the IRAS 100 $\mu$m 
HIRES map of the dust emission, it is more useful to use the higher spatial resolution $(H - K_{s})$-derived extinction 
maps described in $\S$2 to compare the extended dust distribution with the YSO distribution.  Note that the map of A$_{Ks}$ (Fig.~\ref{ak13co_fig_ha}), at the same spatial resolution as the SCUBA map, shows the NE extension from the core indicating higher 
dust density, as seen by SCUBA.  In addition, there are many small regions of enhanced 
extinction, and the extent of the molecular cloud defined by the A$_{Ks}$ map exceeds that suggested 
by the CO maps.  All cloud tracers show an abrupt eastern edge to the cloud, which bifurcates the 
central cluster into the relatively high surface density population embedded within the cloud core, and 
the somewhat lower surface density population outside the cloud.  This edge was likely produced by 
the two HAeBe stars: as stated by \citet{herbig06}, LkH$\alpha$ 324 and 324SE  `have excavated 
a cavity on that edge and cleared out the area to the east, thereby exposing a small cluster of 
H$\alpha$ emission-line stars.'

\section{SPATIAL DISTRIBUTION OF THE YOUNG STELLAR OBJECTS\label{spatial}}

Classification of YSOs was executed as described in $\S$3.  We now examine the spatial 
distribution of the class Is and IIs so identified,  first by overplotting their positions on a 
variety of images: namely, the IR images, the CO image, the cold dust emission images, and 
primarily, the extinction map.

Although the overall {\it stellar} density in the composite NIR image (Fig. 2) is substantially lower 
toward the molecular cloud, the peak surface density of YSOs is found toward the molecular cloud 
core, as identified by all of the molecular cloud tracers.  LkH$\alpha$324 and  LkH$\alpha$324SE 
are associated with that high density region, which we call the `embedded' population of the cluster.   

\begin{figure}\epsscale{.9}
\plotone{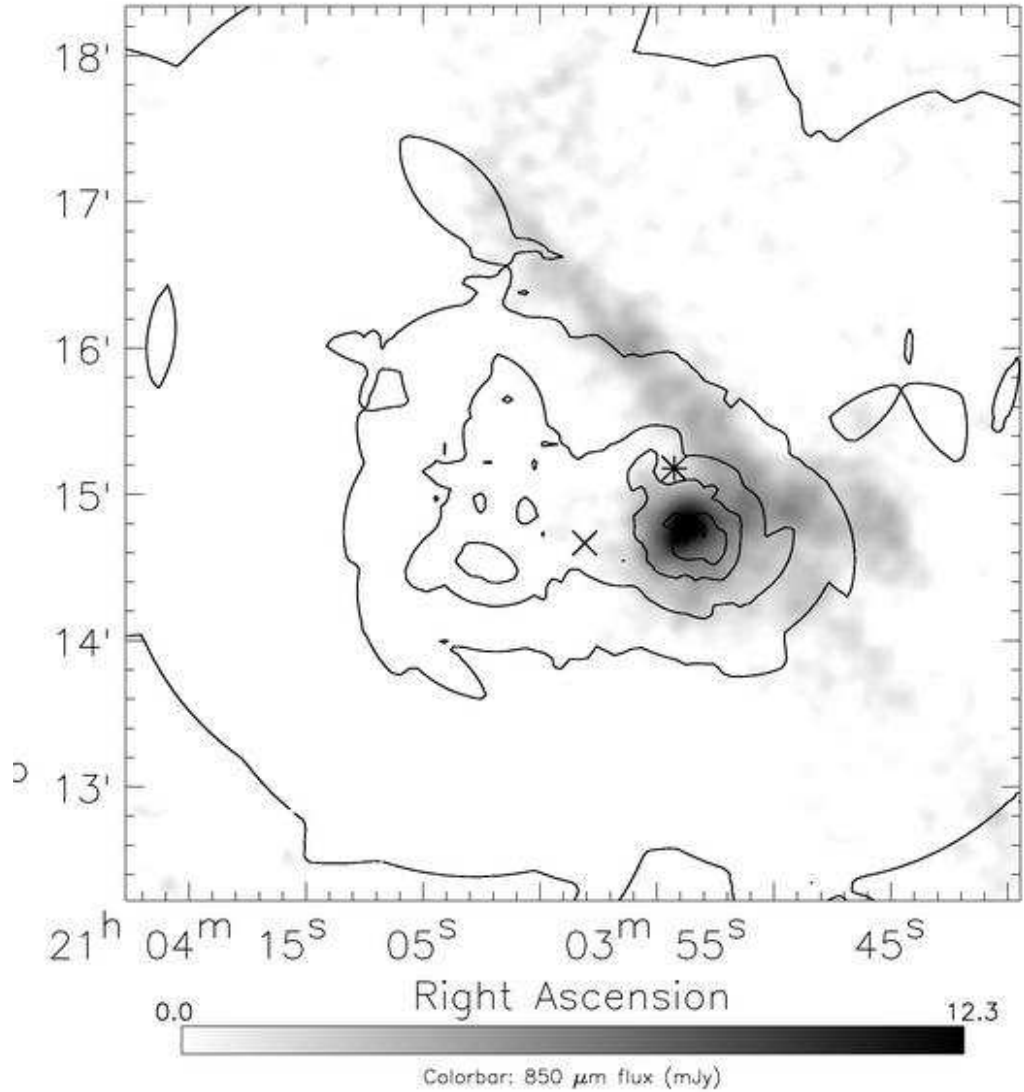}
\caption{Grey scale SCUBA 850 $\mu$m emission in greyscale, overplotted by contours of the 
surface density distributions of classified cluster members.  The locations of LkH$\alpha$ 324 and 
324SE are given by an astrisk and an $\times$, respectively.   The most compact distribution, the 
embedded population, with a relatively high percentage of class Is, is concentrated on the dust 
maximum.  To the east of the dust ridge, there is a second, less compact concentration, of primarily 
class IIs, associated with the emission-line population \citep{herbig06}. 
\label{scubaysocontours_fig}} 
\end{figure}

To illustrate this more clearly, YSO surface density distributions are formed for all YSOs, utilizing 
the method developed for NGC 7129, GGD12-15, and IRAS20050+2720, three clusters previously 
studied in the Young Cluster Project \citep{guter05}.  Because there exists a wide range in stellar 
surface densities over a young cluster, a method which utilizes adaptive smoothing lengths allows 
both the highest and lowest stellar density regions to be probed.  At each sample position [$i,j$] in 
a uniform grid the projected radial distance to the $N$th nearest star, $r_N(i,j)$ is measured, and 
the local surface density $\sigma(i,j) = N/(\pi r^2_N(i,j))$ is calculated.  Here the grid size is 
3$^{\prime\prime}$, and $N$ = 5. A value of  $N \ge 5$ is chosen to retain sufficient spatial 
resolution.  On the other hand \citep{gutermuth05},  $N$ must be as small as feasible to allow small, 
high density sub-groups to be properly represented.  Also see \citet{gutermuth04}. The grid size 
selection of 3$^{\prime\prime}$ was chosen to fully resolve all structure at the  
smallest scales (i.e. highest densities) in the $N$ = 5 density distribution. The surface density of YSOs is 
compared with a SCUBA gray-scale image in Fig.~\ref{scubaysocontours_fig} of the cool dust 
ridge associated with the edge of the molecular cloud. The surface density contours reveal an asymmetric cluster, with the highest surface density region (peaking 
at $\sim$ 1706 stars pc$^{-2}$), including mostly class IIs, concentrated at the peak of the 
dust emission.  LkH$\alpha$ 324 is located on the northern edge of this region 
and LkH$\alpha$ 324 SE lies on the eastern edge.  A second, less compact surface density region associated with the emission-line population (peaking at $\sim$ 747 stars pc$^{-2}$) is 
found to the east of the dust ridge: these are also primarily class II objects (note that class III 
objects are not included in this figure, whereas in the emission-line selected population 
\citep{herbig06} overplotted on the map of A$_{Ks}$ in Fig.~\ref{ak13co_fig_ha}, both class II and III stars are included). 
This population falls in the cavity where the molecular cloud has been excavated, presumably by 
the bright HAeBe stars LkH$\alpha$ 324 and 324 SE. We call this the `exposed' cluster population.

The exposed population of the asymmetric cluster may be somewhat older than the embedded population, as suggested by the 
relatively lower YSO stellar densities encountered, as well as the necessity that enough time has passed 
to sculpt the cloud in the vicinity of the exposed cluster.  We examine this further in the next 
section by estimating disk fractions for the exposed and embedded populations in a circular region 
surrounding the surface density peaks, using standard methods.

In addition to these concentrated distributions of YSOs, in both the excavated cavity, and in the 
dense molecular knot, a distributed halo of YSOs is found mainly in the diffuse portions of the 
associated molecular cloud, and another to the east of the cloud. These haloes include 
predominantly class IIs, but the presence of class Is in the diffuse molecular cloud is 
important to note.  A much larger study focusing on the distributed population of YSOs well 
beyond the clusters, particularly the class Is, is underway. There we hope to determine 
whether the class Is in the distributed population formed in situ, or were ejected from the
clusters.

\section{DISK FRACTIONS IN THE EXPOSED AND EMBEDDED CLUSTER REGIONS\label{fractions}}

We show in this section that $\sim$55$\pm$16\% of the objects in a 1.5$^{\prime}$ diameter circle centered at 
[21$^h$4$^m$4.5$^s$, 50$^{\circ}$14$^{\prime}$32$^{\prime\prime}$.7] are class IIs: this region
corresponds to the peak of the {\it exposed} population of the cluster (here we have purposely not included the 
emission-line objects associated with substantial extinction at the edge of the L988e dark cloud, or 
the extended diffuse halo population, in contrast to the treatment in $\S$4).  This disk fraction, 
while estimated more rigorously than in $\S$4, entails only a relatively small number of stars because L988e is 
sparsely populated. 

In contrast, the disk fraction in a similar sized region centered on the {\it embedded} population's 
peak density (centered at [21$^h$3$^m$52.97$^s$, 50$^\circ$14$^\prime$43$^{\prime\prime}$.5]) is $\sim$74$\pm$15\%.  At face value, the higher mean disk fraction indicates the embedded population is younger than the exposed population, however, because of small number statistics, the large uncertainties overlap, so we cannot be certain of the relative ages of the regions.

\begin{figure}\epsscale{1.}
\plotone{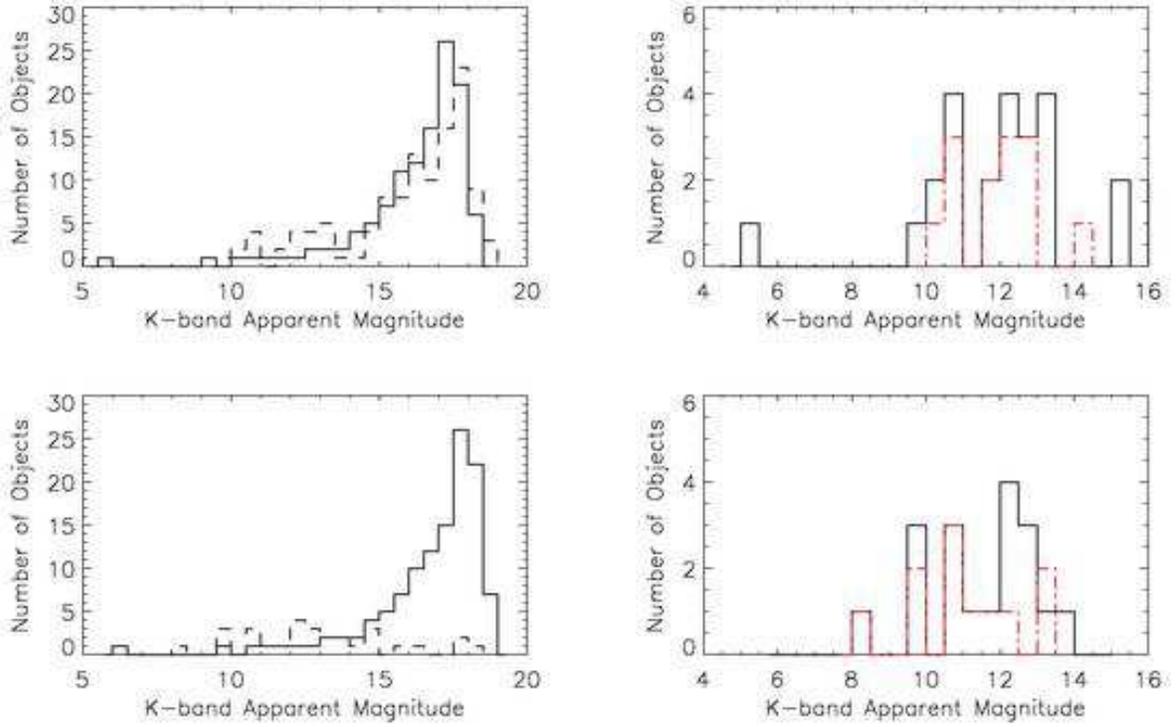}
\caption{Top Left: histograms of $K_{s}$-band detections in 1.5$^{\prime}$ diameter circle centered 
on an off-cloud region (solid line) and on the exposed cluster (dashed line).  Top Right: histograms 
of remaining 22 objects after subtracting the off-cloud sample from the exposed cluster sample 
(black, solid line) and the known population with disks (red, dot-dashed line).  Bottom Left: 
histograms of $K_{s}$-band detections in 1.5$^{\prime}$ diameter circle centered on an off-cloud 
region (solid line) and on the embedded cluster (dashed line).  Bottom Right: histograms of 
remaining 19 objects after subtracting the off-cloud sample from the exposed cluster sample (black, 
solid line) and the known population with disks (red, dot-dashed line). \label{k_hist}} 
\end{figure}

Disk fractions were estimated in the following manner.  $K_{s}$-band magnitude histograms (KMH) 
were created for 1.5$^{\prime}$ diameter regions around the peaks of the exposed and embedded 
populations and for three 3$^{\prime}$ diameter regions off cloud that do not contain any known 
protostars, stars with disks, or H$\alpha$ emission objects (see Fig.~\ref{cover_fig} for the 
locations of these regions and Fig.~\ref{k_hist} for the histograms).  To determine the magnitude at which field star contamination, begins to dominate the cluster 
KMH, we consider the faintest magnitude bin in the cluster KMH greater than 
3$\sigma_{background}$, assuming Poisson counting statistics for each magnitude bin in the 
KMHs.  This limit occurs at 14$^{th}$ magnitude.  The average $K_s$-band extinction 
for the exposed population, the off-cloud regions, and the embedded population are A$_{Ks}$ = 0.37, A$_{Ks}$ 
= 0.28, and A$_{Ks}$ = 1.3, respectively, estimated from $J$,$H$,$K_{s}$, colors as above ($\S$3).  To account for the higher extinction toward the embedded population, the off-cloud sources are artificially extinguished to the mean extinction of the embedded population before 
subtracting the off-cloud KMH from the embedded population KMH.  By subtracting the off-cloud KMH from the exposed population KMH for the magnitude bins less than the 14$^{th}$ magnitude limit, we estimate 
22 cluster members in the exposed population, and 19 cluster members in the 
embedded population: these values are used in the disk fraction determination above.

\section{COMPARISON WITH OTHER CLUSTERS\label{comparison}}

We restrict comparison of L988e to six young clusters \citep{gutermuth05} which have been studied in many of the same ways as L988e, although L988e is the only cluster with exposed stars which have been studied for H$\alpha$ emission.  Of those six, two filamentary clusters, GGD12-15 and IRAS 20050+2720, have the following properties: the class Is follow closely the asymmetric distribution defined by the dust ridges in their respective molecular clouds, while the class IIs are more dispersed along the filament.  The two clusters widely vary in far-IR luminosity and cloud mass, and range from moderate to relatively high spatial density ($\sim$ 1200 and 3000 pc$^{-2}$ peak respectively).  GGD 12-15 has 102 YSOs, of which 25\% are class I, while IRAS 20050+2720 has 121 YSO members, of which 21\% are class I: class I number counts may be biased toward lower numbers due to the lack of MIPS coverage when these regions were analyzed. Two other clusters, NGC 7129 and AFGL 490, are more diffuse than the filamentary clusters, with little high density clustering and little asymmetric structure. There are a total of 63 and 98 YSOs respectively in NGC 7129 and AFGL 490, of which 17\% and 13\% are class I respectively, with the same bias toward lower class I number counts due to lack of MIPS coverage.  YSO spatial distribution in the L988e exposed cluster is similar to that in the latter two clusters, especially to that of NGC 7129: in particular, a large population of class IIs clustered in a cavity in the molecular cloud near a Be star LkH$\alpha$234, resembles the exposed population discussed here.  However, L988e does exhibit some asymmetry, and there is a substantial number of class Is in the L988e embedded cluster population. The distribution of class Is resembles those in GGD12-15 and IRAS 20050+2720: class Is are found in both the high density dusty region near  LkH$\alpha$ 324 and 324-SE, and throughout the rest of the molecular cloud.  However, in contrast to GGD12-15 and IRAS 20050+2720, only 10\% of the 79 class I and II objects identified by IRAC and MIPS in all of L988e are class I.  Of these four clusters, only IRAS 20050+2720 has comparable FIR luminosity to L988e: the other three are $\sim$ 6 - 30 times more luminous.  NGC 7129 is estimated to be $\sim$ 3 Myr in age, while the filamentary clusters are thought to be younger.  The final two young clusters studied in this way, CepA and MonR2, have an order of magnitude higher FIR luminosity, but are otherwise different from each other.  MonR2 exhibits a filamentary distribution of 49 class Is (20\% of the total) similar to that observed for GGD12-15 and IRAS 20050+2720, whereas CepA is sparsely populated by 8 class Is (only 9\% of total YSOs) while the 77 class IIs are anti-associated with CO density enhancements.   A comprehensive spatial distribution study of the entire Young Cluster Survey is in preparation.

\section{SUMMARY}

1.  {\it Spitzer} photometry of the asymmetric L988e cluster in combination with
new NIR photometry reveals the presence of an
embedded population associated with the molecular cloud (the cloud is
defined by CO emission, cool
dust emission, and A$_{Ks}$ maps of extinction). This is in addition to
the previously studied exposed region of the cluster, associated with
the optically studied H$\alpha$ emission-line stars in a cloud cavity on
the eastern edge, sculpted presumably by the HAeBe stars LkH$\alpha$ 324
and LkH$\alpha$ 324 SE.

2.  The disk fraction (class II/(class II + class III) of ONLY the emission-line objects,  
primarily located in the exposed region, is $\sim$ 67\%.  This is consistent with either of the two 
age estimates deduced from $V$ and $I$ photometry \citep{herbig06}. 

3.  The embedded population exhibits larger surface density than the exposed population of the cluster, and is fully enshrouded within L988e's natal 
molecular cloud.  This, as well as the higher $\underline{mean}$ disk fraction calculated in the 
conventional way for all cluster members in the field (74\% vs. 55\%) points to the probability of a 
somewhat younger age for the embedded population as compared with the exposed population, although the uncertainties of these disk fractions overlap.  The 
exposed population of the cluster lies primarily in a region where the extinction A$_{Ks} < $ 0.5 mag.  On the other 
hand, the embedded population suffers a peak extinction of A$_{Ks}$ = 3, and extinctions A$_{Ks} >$ 0.5 throughout the molecular cloud.  

4.  In addition to the asymmetric cluster, there
exists a diffuse halo population of YSOs surrounding both the exposed and
embedded regions of the cluster.  Although the halo is dominated by class
IIs, class Is are also found in the diffuse halo.  Whether these halo objects are
formed in situ is yet to be determined.

5.  Comparing these populations to six regions in the Young Cluster Program studied in a similar 
manner \citep{gutermuth05}, we confirm the conclusion that the spatial 
distribution within the cluster is not strongly coupled to either the FIR luminosity or the numbers of 
cluster members.  The younger clusters more faithfully follow the cloud density structure in which 
the cluster is born, and exhibit higher surface density.  Therefore, L988e is older than the youngest clusters in the survey because the Eastern part of the cloud has had time to be excavated.  It is becoming increasingly obvious that class 
Is are found in the haloes around clusters in regions where the molecular column density is 
relatively low.

6.  {\it Spitzer}/IRAC 3.6$\mu$m, 5.8$\mu$m and {\it Spitzer}/MIPS 24$\mu$m photometry 
reveal another embedded cluster associated with the object L988c.

\acknowledgments

This research has made use of the NASA/IPAC Infrared Science Archive, which is operated by the 
Jet Propulsion Laboratory, California Institute of Technology, under contract with the National 
Aeronautics and Space Administration. This publication makes use of data products from the Two 
Micron All Sky Survey, which is a joint project of the University of Massachusetts and the Infrared 
Processing and Analysis Center/California Institute of Technology, funded by the National 
Aeronautics and Space Administration and the National Science Foundation. JPW acknowledges 
support from NSF grant AST-0324323, and JLP support from SAO/JPL SV4-74011.

\begin{deluxetable}{rccc}
\tabletypesize{\scriptsize}
\tablecaption{L988e Cluster Characteristics Compiled from the Literature.\label{littable}}
\tablewidth{0pt}
\tablehead{
\colhead{ } & \colhead{L988e} 
}
\startdata
Cluster Distance (pc): & 700\tablenotemark{a} \\
Molecular Cloud Mass\tablenotemark{b} ($M_{\odot}$): & 197 \\
Velocity Dispersion\tablenotemark{c} ($km~s^{-1}$): & 1.85(0.79)\\
Far IR Luminosity\tablenotemark{b} ($L_{\odot}$): & 206.1 \\
\enddata
\tablenotetext{a}{\citep{clark86}}
\tablenotetext{b}{C$^{18}$O core mass \citep{ridge03}}
\tablenotetext{c}{Value in parentheses is converted from C$^{18}$O FWHM line width values for 
the core \citep{ridge03} to equivalent Maxwellian values.}
\end{deluxetable}

\clearpage

\begin{deluxetable}{llccccccccccccc}
\rotate
\tabletypesize{\scriptsize}
\setlength{\tabcolsep}{0.02in}
\hspace*{-1in}
\tablecolumns{14}
\tablewidth{0pc}
\tablecaption{Objects with Mid Infrared Excesses (extragalactic contaminants removed): Class I  \label{classItable}}
\tablehead{
\colhead{RA$_{2000}$} & \colhead{DEC$_{2000}$} & \colhead{$J$}\tablenotemark{a} & \colhead{$H$}\tablenotemark{a} & \colhead{$K$}\tablenotemark{a} & \colhead{$[3.6]$}\tablenotemark{a} & \colhead{$[4.5]$}\tablenotemark{a} & \colhead{$[5.8]$}\tablenotemark{a} & \colhead{$[8.0]$}\tablenotemark{a} & \colhead{$[24]$}\tablenotemark{a} & \colhead{A$_{Ks}$}\tablenotemark{b}  & \colhead{$\alpha_{IRAC}$}\tablenotemark{c} & \colhead{IH$\alpha$}\tablenotemark{d} & \colhead{EW(H$\alpha$)}\tablenotemark{e} 
}

\startdata
21:02:50.50$\pm$0.23 &  50:18:52.43$\pm$0.06 &  \nodata &  \nodata &  \nodata &  10.50$\pm$0.003 &  \nodata &  9.66$\pm$0.004 &  \nodata &  5.44$\pm$0.011 &  \nodata &  -0.41\tablenotemark{g} & \nodata & \nodata \\ 
21:02:52.00$\pm$0.32 &  50:13:43.96$\pm$0.19 &  \nodata &  \nodata &  \nodata &  10.59$\pm$0.004 &  \nodata &  9.48$\pm$0.004 &  \nodata &  5.59$\pm$0.014 &  \nodata &  -0.48\tablenotemark{g} &  623 &  26. \\ 
21:02:52.28$\pm$0.05 &  50:12:29.47$\pm$0.12 &  \nodata &  \nodata &  \nodata &  8.55$\pm$0.002 &  \nodata &  7.24$\pm$0.002 &  \nodata &  2.55$\pm$0.003 &  \nodata &  0.020\tablenotemark{g} & \nodata & \nodata \\ 
21:02:54.17$\pm$0.25 &  50:12:27.46$\pm$0.10 &  \nodata &  \nodata &  \nodata &  12.49$\pm$0.007 &  \nodata &  10.01$\pm$0.005 &  \nodata &  4.36$\pm$0.008 &  \nodata &  0.960\tablenotemark{g} & \nodata & \nodata \\ 
21:02:56.60$\pm$0.14 &  50:11:59.22$\pm$0.04 &  \nodata &  \nodata &  \nodata &  10.55$\pm$0.005 &  \nodata &  9.65$\pm$0.007 &  \nodata &  5.40$\pm$0.014 &  \nodata &  -0.37\tablenotemark{g} & \nodata & \nodata \\ 
21:02:58.06$\pm$0.19 &  50:15:25.63$\pm$0.17 &  \nodata &  \nodata &  \nodata &  9.91$\pm$0.002 &  \nodata &  7.82$\pm$0.002 &  \nodata &  3.57$\pm$0.004 &  \nodata &  0.080\tablenotemark{g} & \nodata & \nodata \\ 
21:03:00.19$\pm$0.25 &  50:21:56.59$\pm$0.13 &  \nodata &  \nodata &  \nodata &  11.80$\pm$0.004 &  \nodata &  9.56$\pm$0.004 &  \nodata &  4.03$\pm$0.006 &  \nodata &  0.800\tablenotemark{g} & \nodata & \nodata \\ 
21:03:01.16$\pm$0.01 &  50:12:29.25$\pm$0.03 &  \nodata &  \nodata &  \nodata &  11.60$\pm$0.004 &  \nodata &  10.19$\pm$0.005 &  \nodata &  5.67$\pm$0.018 &  \nodata &  -0.04\tablenotemark{g} & \nodata & \nodata \\ 
21:03:02.98$\pm$0.15 &  50:12:16.36$\pm$0.18 &  \nodata &  \nodata &  \nodata &  12.60$\pm$0.007 &  \nodata &  11.59$\pm$0.014 &  \nodata &  7.62$\pm$0.076 &  \nodata &  -0.47\tablenotemark{g} & \nodata & \nodata \\ 
21:03:03.24$\pm$0.12 &  50:13:12.66$\pm$0.06 &  \nodata &  \nodata &  \nodata &  9.79$\pm$0.004 &  \nodata &  8.60$\pm$0.041 &  \nodata &  2.95$\pm$0.014 &  \nodata &  0.470\tablenotemark{g} & \nodata & \nodata \\ 
21:03:03.34$\pm$0.43 &  50:09:31.94$\pm$0.15 &  \nodata &  \nodata &  \nodata &  12.92$\pm$0.007 &  12.26$\pm$0.006 &  11.61$\pm$0.012 &  10.48$\pm$0.011 & \nodata &  \nodata &  -0.05 & \nodata & \nodata \\ 
21:03:06.85$\pm$0.17 &  50:12:05.20$\pm$0.37 &  \nodata &  \nodata &  \nodata &  11.77$\pm$0.005 &  11.29$\pm$0.004 &  10.60$\pm$0.008 &  9.54$\pm$0.005 &  6.72$\pm$0.022 &  \nodata &  -0.25 & \nodata & \nodata \\ 
21:03:13.90$\pm$0.21 &  50:12:46.78$\pm$0.06 &  \nodata &  \nodata &  \nodata &  11.37$\pm$0.004 &  10.57$\pm$0.004 &  10.03$\pm$0.006 &  9.20$\pm$0.005 &  6.16$\pm$0.015 &  \nodata &  -0.40 & \nodata & \nodata \\ 
21:03:22.92$\pm$0.22 &  50:11:31.25$\pm$0.23 &  \nodata &  \nodata &  \nodata &  16.33$\pm$0.066 &  15.19$\pm$0.052 &  14.06$\pm$0.063 &  12.96$\pm$0.067 &  9.60$\pm$0.151 &  \nodata &  1.040 & \nodata & \nodata \\ 
21:03:35.66$\pm$0.24 &  50:15:25.77$\pm$0.33 &  \nodata &  16.57$\pm$0.066 &  15.59$\pm$0.039 &  13.92$\pm$0.030 &  13.28$\pm$0.017 &  12.92$\pm$0.096 &  11.93$\pm$0.049 &  6.52$\pm$0.026 &  1.09 &  -0.64 & \nodata & \nodata \\ 
21:03:53.44$\pm$0.42 &  50:14:47.55$\pm$0.39 &  \nodata &  \nodata &  \nodata &  \nodata &  8.68$\pm$0.008 &  7.00$\pm$0.018 &  5.81$\pm$0.028 &  1.35$\pm$0.042 &  \nodata & 0.98\tablenotemark{f} & \nodata & \nodata \\ 
21:03:56.37$\pm$0.02 &  50:16:03.27$\pm$0.08 &  \nodata &  \nodata &  \nodata &  13.59$\pm$0.013 &  11.25$\pm$0.005 &  9.90$\pm$0.007 &  8.62$\pm$0.007 &  3.97$\pm$0.024 &  \nodata &  2.720 & \nodata & \nodata \\ 
21:03:58.13$\pm$0.34 &  50:14:39.77$\pm$0.50 &  7.71$\pm$0.015 &  6.51$\pm$0.003 &  5.32$\pm$0.011 &  4.42$\pm$0.006 &  3.64$\pm$0.005 &  1.96$\pm$0.001 &  1.65$\pm$0.001 & \nodata &  1.14 &  0.580 &  651 &  159 \\ 
21:04:00.97$\pm$0.48 &  50:15:56.77$\pm$0.08 &  14.09$\pm$0.005 &  13.08$\pm$0.002 &  12.23$\pm$0.001 &  10.25$\pm$0.002 &  9.50$\pm$0.003 &  8.79$\pm$0.009 &  7.79$\pm$0.005 &  4.13$\pm$0.027 &  0.68 &  -0.03 &  657 &  69. \\

\enddata
\tablenotetext{a}{Magnitudes are de-reddened when extinction value A$_{Ks}$ quoted.}
\tablenotetext{b}{A$_K$ determined from method outlined in $\S$5.}
\tablenotetext{c}{Slope $\alpha$ determined from IRAC photometry}
\tablenotetext{d}{Running number from \citet{herbig06}.}
\tablenotetext{e}{ EW[H$\alpha$] from \citet{herbig06}.}
\tablenotetext{g}{Slope $\alpha$ determined from [3.6], [5.8] and [24] photometry}
\tablenotetext{f}{Slope $\alpha$ determined from [4.5] to [24] photometry}
\end{deluxetable}



\begin{deluxetable}{llccccccccccccc}
\tabletypesize{\scriptsize}
\setlength{\tabcolsep}{0.02in}
\tablecolumns{14}
\tablewidth{0pc}
\rotate
\tablecaption{Objects with Mid Infrared Excesses (extragalactic contaminants removed): Class II
\label{classIItable}}
\tablehead{
\colhead{RA$_{2000}$} & \colhead{DEC$_{2000}$} & \colhead{$J$}\tablenotemark{a} & \colhead{$H$}\tablenotemark{a} & \colhead{$K$}\tablenotemark{a} & \colhead{$[3.6]$}\tablenotemark{a} & \colhead{$[4.5]$}\tablenotemark{a} & \colhead{$[5.8]$}\tablenotemark{a} & \colhead{$[8.0]$}\tablenotemark{a} & \colhead{$[24]$}\tablenotemark{a} & \colhead{A$_{Ks}$}\tablenotemark{b}  & \colhead{$\alpha_{IRAC}$}\tablenotemark{c} & \colhead{IH$\alpha$}\tablenotemark{d} & \colhead{EW(H$\alpha$)}\tablenotemark{e} 
}

\startdata
21:02:55.21$\pm$0.05 &  50:11:55.22$\pm$0.22 &  \nodata &  \nodata &  \nodata &  9.06$\pm$0.006 &  \nodata &  7.68$\pm$0.004 &  \nodata &  4.34$\pm$0.008 &  \nodata &  -0.66\tablenotemark{f} &  624 &  29. \\ 
21:02:59.28$\pm$0.03 &  50:10:30.24$\pm$0.13 &  \nodata &  \nodata &  \nodata &  11.61$\pm$0.004 &  \nodata &  10.70$\pm$0.007 &  \nodata &  6.94$\pm$0.061 &  \nodata &  -0.62\tablenotemark{f} & \nodata & \nodata \\ 
21:03:06.34$\pm$0.24 &  50:10:04.04$\pm$0.08 &  \nodata &  \nodata &  \nodata &  11.06$\pm$0.004 &  10.48$\pm$0.004 &  9.99$\pm$0.005 &  9.59$\pm$0.005 &  7.87$\pm$0.047 &  \nodata &  -1.16 & \nodata & \nodata \\ 
21:03:08.09$\pm$0.15 &  50:10:01.71$\pm$0.20 &  \nodata &  \nodata &  \nodata &  14.31$\pm$0.020 &  13.92$\pm$0.029 &  13.57$\pm$0.062 &  13.39$\pm$0.093 &  \nodata &  \nodata &  -1.78 & \nodata & \nodata \\ 
21:03:20.22$\pm$0.49 &  50:15:58.91$\pm$0.43 &  9.94$\pm$0.008 &  9.34$\pm$0.001 &  9.22$\pm$0.001 &  8.49$\pm$0.002 &  8.14$\pm$0.003 &  7.60$\pm$0.003 &  6.64$\pm$0.002 &  3.55$\pm$0.005 &  2.15 &  -0.70 & \nodata & \nodata \\ 
21:03:21.78$\pm$0.27 &  50:09:41.73$\pm$0.10 &  15.06$\pm$0.011 &  14.36$\pm$0.005 &  14.05$\pm$0.007 &  13.38$\pm$0.013 &  13.18$\pm$0.029 &  13.16$\pm$0.053 &  13.17$\pm$0.070 &  \nodata &  0.53 &  -2.64 & \nodata & \nodata \\ 
21:03:23.72$\pm$0.17 &  50:17:21.65$\pm$0.56 &  10.24$\pm$0.088 &  9.60$\pm$0.004 &  9.40$\pm$0.001 &  9.09$\pm$0.003 &  8.73$\pm$0.003 &  8.29$\pm$0.005 &  7.72$\pm$0.004 &  5.94$\pm$0.021 &  3.56 &  -1.25 & \nodata & \nodata \\ 
21:03:24.17$\pm$0.32 &  50:11:01.01$\pm$0.34 &  \nodata &  \nodata &  \nodata &  12.81$\pm$0.006 &  12.11$\pm$0.005 &  11.65$\pm$0.015 &  11.77$\pm$0.061 &  \nodata &  \nodata &  -1.65 & \nodata & \nodata \\ 
21:03:26.47$\pm$0.37 &  50:14:28.25$\pm$0.60 &  15.06$\pm$0.012 &  14.22$\pm$0.005 &  13.68$\pm$0.004 &  12.79$\pm$0.008 &  12.44$\pm$0.008 &  12.03$\pm$0.020 &  11.52$\pm$0.026 &  7.76$\pm$0.043 &  0.87 &  -1.37 & \nodata & \nodata \\ 
21:03:28.67$\pm$0.78 &  50:12:34.94$\pm$0.47 &  14.26$\pm$0.064 &  13.36$\pm$0.009 &  12.70$\pm$0.005 &  12.31$\pm$0.017 &  12.16$\pm$0.015 &  12.06$\pm$0.037 &  12.13$\pm$0.088 &  \nodata &  1.81 &  -2.64 & \nodata & \nodata \\ 
21:03:30.74$\pm$0.70 &  50:12:49.04$\pm$0.54 &  12.52$\pm$0.046 &  11.80$\pm$0.004 &  11.46$\pm$0.002 &  11.00$\pm$0.005 &  10.73$\pm$0.005 &  10.22$\pm$0.010 &  9.66$\pm$0.008 &  7.45$\pm$0.051 &  2.31 &  -1.26 & \nodata & \nodata \\ 
21:03:36.20$\pm$0.48 &  50:15:56.67$\pm$0.33 &  13.24$\pm$0.005 &  12.50$\pm$0.002 &  12.12$\pm$0.001 &  11.38$\pm$0.004 &  10.98$\pm$0.004 &  10.51$\pm$0.009 &  9.92$\pm$0.010 &  7.83$\pm$0.072 &  1.07 &  -1.14 & \nodata & \nodata \\ 
21:03:36.24$\pm$0.81 &  50:11:50.14$\pm$0.53 &  16.57$\pm$0.035 &  15.72$\pm$0.016 &  15.15$\pm$0.015 &  14.25$\pm$0.012 &  13.89$\pm$0.015 &  13.41$\pm$0.052 &  12.56$\pm$0.064 &  \nodata &  0.47 &  -0.90 & \nodata & \nodata \\ 
21:03:36.69$\pm$0.45 &  50:16:49.22$\pm$0.13 &  14.58$\pm$0.035 &  13.76$\pm$0.007 &  13.23$\pm$0.005 &  12.36$\pm$0.007 &  12.12$\pm$0.007 &  11.65$\pm$0.018 &  11.12$\pm$0.022 &  8.30$\pm$0.080 &  1.40 &  -1.38 & \nodata & \nodata \\ 
21:03:37.44$\pm$0.56 &  50:13:36.68$\pm$0.37 &  \nodata &  13.99$\pm$0.028 &  13.69$\pm$0.015 &  13.24$\pm$0.017 &  13.12$\pm$0.015 &  12.73$\pm$0.066 &  12.56$\pm$0.081 &  \nodata &  2.31 &  -2.01 & \nodata & \nodata \\ 
21:03:39.47$\pm$0.32 &  50:15:52.84$\pm$0.27 &  12.40$\pm$0.001 &  11.67$\pm$0.019 &  11.30$\pm$0.022 &  10.69$\pm$0.002 &  10.52$\pm$0.003 &  10.07$\pm$0.006 &  9.48$\pm$0.008 &  6.78$\pm$0.059 &  0.14 &  -1.41 &  631 &  55. \\ 
21:03:39.68$\pm$0.21 &  50:12:36.28$\pm$0.55 &  \nodata &  12.40$\pm$0.028 &  11.98$\pm$0.009 &  11.50$\pm$0.008 &  11.28$\pm$0.009 &  10.78$\pm$0.024 &  10.43$\pm$0.054 &  \nodata &  3.26 &  -1.55 & \nodata & \nodata \\ 
21:03:40.70$\pm$0.11 &  50:13:54.05$\pm$0.05 &  11.19$\pm$0.001 &  10.54$\pm$0.001 &  10.31$\pm$0.022 &  9.73$\pm$0.003 &  9.48$\pm$0.003 &  9.09$\pm$0.004 &  8.57$\pm$0.004 &  6.03$\pm$0.026 &  1.18 &  -1.49 & \nodata & \nodata \\ 
21:03:43.58$\pm$0.27 &  50:15:58.23$\pm$0.00 &  11.01$\pm$0.001 &  10.16$\pm$0.017 &  9.60$\pm$0.019 &  8.25$\pm$0.002 &  7.82$\pm$0.002 &  7.35$\pm$0.002 &  6.97$\pm$0.002 &  4.25$\pm$0.012 &  0.77 &  -1.36 &  632 &  39. \\ 
21:03:44.23$\pm$0.18 &  50:16:05.00$\pm$0.53 &  14.52$\pm$0.030 &  13.92$\pm$0.008 &  13.92$\pm$0.010 &  13.17$\pm$0.039 &  12.98$\pm$0.066 &  \nodata &  \nodata &  \nodata &  1.46 & \nodata & \nodata & \nodata \\ 
21:03:44.46$\pm$0.34 &  50:15:51.63$\pm$0.33 &  12.28$\pm$0.002 &  11.60$\pm$0.001 &  11.31$\pm$0.001 &  10.66$\pm$0.009 &  10.23$\pm$0.007 &  9.81$\pm$0.010 &  9.22$\pm$0.009 &  \nodata &  0.99 &  -1.19 &  633 &  156 \\ 
21:03:46.39$\pm$0.24 &  50:13:44.79$\pm$0.05 &  15.61$\pm$0.003 &  14.77$\pm$0.002 &  14.22$\pm$0.002 &  13.50$\pm$0.015 &  13.15$\pm$0.018 &  12.81$\pm$0.050 &  12.35$\pm$0.084 &  7.85$\pm$0.155 &  0.09 &  -1.52 & \nodata & \nodata \\ 
21:03:46.58$\pm$0.27 &  50:08:41.97$\pm$0.82 &  12.51$\pm$0.001 &  11.85$\pm$0.001 &  11.60$\pm$0.001 &  11.32$\pm$0.004 &  11.15$\pm$0.004 &  10.87$\pm$0.012 &  10.25$\pm$0.036 &  7.57$\pm$0.115 &  0.15 &  -1.62 &  635 &  4.2 \\ 
21:03:46.71$\pm$0.43 &  50:15:26.33$\pm$0.24 &  \nodata &  13.22$\pm$0.009 &  12.33$\pm$0.004 &  11.01$\pm$0.005 &  10.43$\pm$0.004 &  9.72$\pm$0.014 &  8.98$\pm$0.035 &  6.47$\pm$0.186 &  1.87 &  -0.49 & \nodata & \nodata \\ 
21:03:47.19$\pm$0.20 &  50:11:27.63$\pm$0.29 &  13.25$\pm$0.001 &  12.62$\pm$0.001 &  12.44$\pm$0.001 &  12.31$\pm$0.006 &  11.94$\pm$0.005 &  11.61$\pm$0.015 &  10.73$\pm$0.015 &  8.02$\pm$0.116 &  0.19 &  -1.06 &  637 &  19. \\ 
21:03:48.42$\pm$0.37 &  50:14:18.15$\pm$0.13 &  11.08$\pm$0.011 &  10.29$\pm$0.001 &  9.84$\pm$0.001 &  9.11$\pm$0.003 &  8.68$\pm$0.003 &  8.15$\pm$0.004 &  7.44$\pm$0.003 &  3.32$\pm$0.006 &  1.97 &  -0.91 & \nodata & \nodata \\ 
21:03:48.89$\pm$0.12 &  50:10:04.72$\pm$0.54 &  13.33$\pm$0.001 &  12.64$\pm$0.001 &  12.33$\pm$0.001 &  11.96$\pm$0.005 &  11.78$\pm$0.005 &  11.65$\pm$0.019 &  10.94$\pm$0.055 &  6.20$\pm$0.043 &  0.27 &  -1.71 &  638 &  28. \\ 
21:03:49.03$\pm$0.14 &  50:16:02.59$\pm$0.26 &  12.13$\pm$0.002 &  11.25$\pm$0.001 &  10.62$\pm$0.001 &  9.55$\pm$0.002 &  9.23$\pm$0.003 &  8.97$\pm$0.004 &  8.56$\pm$0.016 &  6.26$\pm$0.064 &  0.95 &  -1.72 &  639 &  6.8 \\ 
21:03:50.14$\pm$0.01 &  50:10:23.79$\pm$0.06 &  13.26$\pm$0.001 &  12.45$\pm$0.001 &  11.95$\pm$0.001 &  11.26$\pm$0.004 &  10.85$\pm$0.003 &  10.48$\pm$0.007 &  9.85$\pm$0.010 &  8.10$\pm$0.286 &  0.22 &  -1.24 &  640 &  86. \\ 
21:03:50.80$\pm$0.08 &  50:15:33.02$\pm$0.19 &  14.12$\pm$0.077 &  13.35$\pm$0.011 &  12.92$\pm$0.006 &  12.68$\pm$0.047 &  12.50$\pm$0.040 &  \nodata &  \nodata &  \nodata &  1.94 & \nodata & \nodata & \nodata \\ 
21:03:51.50$\pm$0.11 &  50:14:32.72$\pm$0.31 &  10.81$\pm$0.103 &  10.18$\pm$0.006 &  9.62$\pm$0.001 &  8.68$\pm$0.005 &  8.37$\pm$0.004 &  7.99$\pm$0.016 &  7.63$\pm$0.041 &  3.95$\pm$0.021 &  3.29 &  -1.62 & \nodata & \nodata \\ 
21:03:52.17$\pm$0.02 &  50:14:43.89$\pm$0.34 &  12.40$\pm$0.005 &  11.30$\pm$0.001 &  10.30$\pm$0.001 &  9.12$\pm$0.007 &  8.58$\pm$0.006 &  8.03$\pm$0.026 &  7.54$\pm$0.059 &  \nodata &  1.33 &  -1.00 & \nodata & \nodata \\ 
21:03:52.89$\pm$0.05 &  50:14:38.07$\pm$0.09 &  14.42$\pm$0.004 &  13.64$\pm$0.003 &  13.18$\pm$0.004 &  12.51$\pm$0.070 &  12.06$\pm$0.078 &  \nodata &  \nodata &  \nodata &  0.25 & \nodata & \nodata & \nodata \\ 
21:03:53.32$\pm$0.31 &  50:17:49.99$\pm$0.32 &  16.96$\pm$0.030 &  16.20$\pm$0.018 &  15.80$\pm$0.020 &  14.66$\pm$0.034 &  14.52$\pm$0.028 &  13.88$\pm$0.078 &  13.32$\pm$0.090 &  \nodata &  0.44 &  -1.21 & \nodata & \nodata \\ 
21:03:53.56$\pm$0.31 &  50:16:00.64$\pm$0.07 &  14.05$\pm$0.003 &  13.24$\pm$0.002 &  12.75$\pm$0.001 &  12.07$\pm$0.006 &  11.69$\pm$0.008 &  11.51$\pm$0.045 &  10.80$\pm$0.106 &  \nodata &  0.47 &  -1.44 & \nodata & \nodata \\ 
21:03:53.87$\pm$0.01 &  50:14:44.48$\pm$0.01 &  11.50$\pm$0.006 &  10.77$\pm$0.002 &  10.41$\pm$0.001 &  9.86$\pm$0.034 &  9.49$\pm$0.029 &  \nodata &  \nodata &  \nodata &  1.44 & \nodata & \nodata & \nodata \\ 
21:03:53.91$\pm$0.00 &  50:15:14.39$\pm$0.23 &  14.12$\pm$0.005 &  13.32$\pm$0.004 &  12.84$\pm$0.004 &  11.63$\pm$0.062 &  11.10$\pm$0.051 &  \nodata &  \nodata &  \nodata &  0.28 & \nodata &  643 &  43. \\ 
21:03:54.10$\pm$0.05 &  50:15:29.26$\pm$0.11 &  12.24$\pm$0.002 &  11.53$\pm$0.001 &  11.21$\pm$0.001 &  10.49$\pm$0.009 &  10.00$\pm$0.007 &  9.65$\pm$0.074 &  8.70$\pm$0.158 &  \nodata &  1.06 &  -0.84 & \nodata & \nodata \\ 
21:03:54.21$\pm$0.32 &  50:15:10.15$\pm$0.13 &  10.02$\pm$10.00 &  9.81$\pm$0.015 &  9.57$\pm$0.090 &  9.51$\pm$0.017 &  9.27$\pm$0.018 &  8.50$\pm$0.109 &  \nodata &  \nodata &  0.00 & \nodata & \nodata &  11. \\ 
21:03:54.36$\pm$0.16 &  50:15:00.07$\pm$0.01 &  12.57$\pm$0.002 &  11.94$\pm$0.001 &  11.74$\pm$0.001 &  11.00$\pm$0.094 &  10.64$\pm$0.056 &  \nodata &  \nodata &  \nodata &  0.45 & \nodata &  644 &  15. \\ 
21:03:54.43$\pm$0.37 &  50:13:50.73$\pm$0.28 &  11.90$\pm$0.003 &  11.26$\pm$0.001 &  11.04$\pm$0.001 &  10.47$\pm$0.004 &  10.30$\pm$0.006 &  10.10$\pm$0.019 &  9.79$\pm$0.032 &  \nodata &  1.47 &  -2.06 & \nodata & \nodata \\ 
21:03:54.63$\pm$0.11 &  50:14:36.87$\pm$0.25 &  8.79$\pm$0.025 &  7.89$\pm$0.021 &  7.23$\pm$0.037 &  5.88$\pm$0.002 &  5.25$\pm$0.002 &  4.51$\pm$0.001 &  3.56$\pm$0.003 &  0.95$\pm$0.006 &  1.49 &  -0.16 &  645 & \nodata \\ 
21:03:55.05$\pm$0.30 &  50:14:58.53$\pm$0.12 &  12.84$\pm$0.007 &  11.97$\pm$0.003 &  11.35$\pm$0.002 &  10.09$\pm$0.046 &  9.60$\pm$0.023 &  \nodata &  \nodata &  \nodata &  1.24 & \nodata & \nodata & \nodata \\ 
21:03:55.63$\pm$0.04 &  50:14:14.28$\pm$0.09 &  15.21$\pm$0.024 &  14.26$\pm$0.008 &  13.51$\pm$0.006 &  12.18$\pm$0.044 &  11.75$\pm$0.071 &  11.45$\pm$0.165 &  10.14$\pm$0.142 &  \nodata &  0.95 &  -0.57 & \nodata & \nodata \\ 
21:03:56.11$\pm$0.01 &  50:10:02.70$\pm$0.14 &  13.78$\pm$0.002 &  12.91$\pm$0.001 &  12.31$\pm$0.001 &  11.50$\pm$0.005 &  11.24$\pm$0.004 &  10.88$\pm$0.014 &  10.69$\pm$0.047 &  \nodata &  0.00 &  -1.89 &  647 &  19. \\ 
21:03:56.33$\pm$0.14 &  50:14:51.39$\pm$0.20 &  12.00$\pm$0.001 &  11.21$\pm$0.001 &  10.73$\pm$0.001 &  9.53$\pm$0.007 &  9.15$\pm$0.012 &  8.58$\pm$0.038 &  8.79$\pm$0.285 &  \nodata &  0.24 & \nodata &  648 &  31. \\ 
21:03:56.42$\pm$0.27 &  50:10:03.46$\pm$0.32 &  12.88$\pm$0.001 &  12.22$\pm$0.001 &  11.98$\pm$0.001 &  11.77$\pm$0.006 &  11.49$\pm$0.005 &  \nodata &  \nodata &  \nodata &  0.06 & \nodata & \nodata & \nodata \\ 
21:03:56.95$\pm$0.57 &  50:13:51.61$\pm$0.28 &  \nodata &  13.72$\pm$0.045 &  13.47$\pm$0.020 &  12.90$\pm$0.046 &  12.82$\pm$0.087 &  \nodata &  \nodata &  \nodata &  2.75 & \nodata & \nodata & \nodata \\ 
21:03:56.96$\pm$0.08 &  50:15:17.71$\pm$0.13 &  13.40$\pm$0.001 &  12.67$\pm$0.001 &  12.30$\pm$0.001 &  11.45$\pm$0.017 &  10.84$\pm$0.015 &  10.72$\pm$0.166 &  10.12$\pm$0.562 &  \nodata &  0.35 & \nodata &  649 &  233 \\ 
21:03:58.90$\pm$0.13 &  50:14:50.88$\pm$0.34 &  11.33$\pm$0.001 &  10.55$\pm$0.024 &  10.12$\pm$0.001 &  9.06$\pm$0.011 &  8.54$\pm$0.029 &  8.25$\pm$0.072 &  7.82$\pm$0.157 &  \nodata &  0.33 &  -1.47 &  652 &  53. \\ 
21:03:59.24$\pm$0.37 &  50:14:57.66$\pm$0.15 &  11.16$\pm$0.001 &  10.48$\pm$0.001 &  10.21$\pm$0.001 &  9.47$\pm$0.012 &  9.13$\pm$0.017 &  8.60$\pm$0.134 &  7.91$\pm$0.132 &  \nodata &  0.75 &  -1.02 &  654 &  53. \\ 
21:04:00.18$\pm$0.36 &  50:15:24.36$\pm$0.09 &  12.47$\pm$0.001 &  11.70$\pm$0.017 &  11.28$\pm$0.022 &  10.45$\pm$0.003 &  10.02$\pm$0.003 &  9.54$\pm$0.028 &  8.82$\pm$0.040 &  \nodata &  0.11 &  -0.96 &  655 &  98. \\ 
21:04:00.94$\pm$0.17 &  50:17:08.63$\pm$0.36 &  13.77$\pm$0.019 &  13.06$\pm$0.005 &  12.72$\pm$0.004 &  12.44$\pm$0.011 &  12.20$\pm$0.013 &  12.16$\pm$0.044 &  12.50$\pm$0.125 &  \nodata &  1.39 &  -2.93 & \nodata & \nodata \\ 
21:04:00.95$\pm$0.20 &  50:17:21.39$\pm$0.17 &  \nodata &  12.36$\pm$0.011 &  12.08$\pm$0.007 &  11.66$\pm$0.039 &  11.55$\pm$0.042 &  10.62$\pm$0.053 &  \nodata &  \nodata &  2.40 & \nodata & \nodata & \nodata \\ 
21:04:01.02$\pm$0.02 &  50:17:27.05$\pm$0.08 &  10.94$\pm$0.002 &  9.95$\pm$0.001 &  9.16$\pm$0.021 &  8.12$\pm$0.003 &  7.64$\pm$0.002 &  7.18$\pm$0.002 &  6.64$\pm$0.002 &  4.53$\pm$0.008 &  1.58 &  -1.15 & \nodata & \nodata \\ 
21:04:01.34$\pm$0.32 &  50:14:41.85$\pm$0.30 &  12.63$\pm$0.001 &  11.74$\pm$0.035 &  11.01$\pm$0.031 &  9.81$\pm$0.004 &  9.42$\pm$0.005 &  8.80$\pm$0.032 &  8.28$\pm$0.041 &  \nodata &  0.00 &  -1.03 &  658 &  4.1 \\ 
21:04:01.50$\pm$0.08 &  50:15:05.13$\pm$0.25 &  13.52$\pm$0.002 &  12.75$\pm$0.001 &  12.30$\pm$0.001 &  11.51$\pm$0.010 &  11.08$\pm$0.020 &  10.74$\pm$0.052 &  9.86$\pm$0.076 &  \nodata &  0.23 &  -0.99 &  659 &  183 \\ 
21:04:01.77$\pm$0.11 &  50:14:23.22$\pm$0.38 &  13.44$\pm$0.001 &  12.76$\pm$0.001 &  12.47$\pm$0.001 &  12.24$\pm$0.029 &  12.05$\pm$0.036 &  11.76$\pm$0.267 &  \nodata &  \nodata &  0.27 & \nodata & \nodata & \nodata \\ 
21:04:01.87$\pm$0.53 &  50:14:43.11$\pm$0.27 &  13.21$\pm$0.001 &  12.40$\pm$0.001 &  11.89$\pm$0.001 &  11.25$\pm$0.012 &  10.82$\pm$0.014 &  10.29$\pm$0.077 &  9.28$\pm$0.051 &  \nodata &  0.26 &  -0.58 &  660 &  121 \\ 
21:04:02.03$\pm$0.24 &  50:14:10.10$\pm$0.27 &  14.92$\pm$0.003 &  14.20$\pm$0.002 &  13.86$\pm$0.003 &  13.45$\pm$0.041 &  13.17$\pm$0.055 &  \nodata &  \nodata &  \nodata &  0.13 & \nodata & \nodata & \nodata \\ 
21:04:02.77$\pm$0.19 &  50:14:20.31$\pm$0.05 &  11.08$\pm$0.013 &  10.35$\pm$0.015 &  9.97$\pm$0.019 &  8.91$\pm$0.003 &  8.36$\pm$0.002 &  7.63$\pm$0.005 &  6.66$\pm$0.010 &  3.27$\pm$0.027 &  0.12 &  -0.23 &  663 &  12. \\ 
21:04:02.80$\pm$0.18 &  50:15:29.56$\pm$0.24 &  12.17$\pm$0.001 &  11.55$\pm$0.001 &  11.38$\pm$0.001 &  11.02$\pm$0.005 &  10.54$\pm$0.004 &  10.00$\pm$0.012 &  9.26$\pm$0.020 &  5.48$\pm$0.127 &  0.65 &  -0.81 &  664 &  9.6 \\ 
21:04:03.36$\pm$0.17 &  50:14:44.47$\pm$0.12 &  13.24$\pm$0.001 &  12.34$\pm$0.021 &  11.69$\pm$0.001 &  10.77$\pm$0.004 &  10.35$\pm$0.004 &  9.83$\pm$0.024 &  9.03$\pm$0.016 &  \nodata &  0.03 &  -0.84 &  666 &  46. \\ 
21:04:05.50$\pm$0.16 &  50:14:36.15$\pm$0.24 &  15.54$\pm$0.003 &  14.81$\pm$0.003 &  14.35$\pm$0.003 &  13.63$\pm$0.028 &  13.16$\pm$0.025 &  12.51$\pm$0.085 &  12.00$\pm$0.046 &  \nodata &  0.00 &  -0.93 & \nodata & \nodata \\ 
21:04:05.72$\pm$0.38 &  50:14:47.85$\pm$0.19 &  13.30$\pm$0.002 &  12.65$\pm$0.001 &  12.44$\pm$0.001 &  11.85$\pm$0.011 &  11.59$\pm$0.007 &  11.19$\pm$0.030 &  10.56$\pm$0.056 &  6.86$\pm$0.162 &  0.28 &  -1.34 & \nodata & \nodata \\ 
21:04:06.11$\pm$0.09 &  50:16:39.01$\pm$0.34 &  12.62$\pm$0.001 &  11.85$\pm$0.001 &  11.42$\pm$0.001 &  10.23$\pm$0.004 &  9.83$\pm$0.003 &  9.46$\pm$0.004 &  8.62$\pm$0.006 &  5.45$\pm$0.017 &  0.53 &  -1.01 &  670 &  28. \\ 
21:04:06.82$\pm$0.09 &  50:13:17.63$\pm$0.34 &  11.33$\pm$0.001 &  10.65$\pm$0.015 &  10.37$\pm$0.021 &  10.11$\pm$0.003 &  9.55$\pm$0.003 &  9.11$\pm$0.004 &  8.25$\pm$0.004 &  5.12$\pm$0.031 &  0.73 &  -0.74 &  671 &  161 \\ 
21:04:06.86$\pm$0.09 &  50:14:41.69$\pm$0.16 &  13.72$\pm$0.001 &  12.93$\pm$0.001 &  12.48$\pm$0.002 &  11.78$\pm$0.009 &  11.43$\pm$0.007 &  11.04$\pm$0.022 &  10.51$\pm$0.060 &  7.44$\pm$0.173 &  0.04 &  -1.38 &  672 &  101 \\ 
21:04:07.05$\pm$0.20 &  50:18:52.32$\pm$0.12 &  12.01$\pm$0.004 &  11.41$\pm$0.001 &  11.29$\pm$0.001 &  10.76$\pm$0.004 &  10.65$\pm$0.004 &  10.15$\pm$0.010 &  9.24$\pm$0.006 &  5.57$\pm$0.013 &  1.32 &  -1.04 & \nodata & \nodata \\ 
21:04:07.09$\pm$0.05 &  50:15:19.01$\pm$0.17 &  12.96$\pm$0.002 &  12.35$\pm$0.002 &  12.20$\pm$0.002 &  11.80$\pm$0.009 &  11.55$\pm$0.012 &  11.18$\pm$0.014 &  10.49$\pm$0.026 &  6.92$\pm$0.059 &  0.41 &  -1.33 &  673 &  17. \\ 
21:04:11.23$\pm$0.03 &  50:10:48.57$\pm$0.07 &  12.03$\pm$0.001 &  11.33$\pm$0.017 &  11.03$\pm$0.021 &  10.21$\pm$0.002 &  9.94$\pm$0.003 &  9.56$\pm$0.004 &  8.94$\pm$0.004 &  6.97$\pm$0.031 &  0.27 &  -1.38 &  676 &  26. \\ 
21:04:19.50$\pm$0.24 &  50:15:57.86$\pm$0.22 &  11.88$\pm$0.001 &  11.00$\pm$0.001 &  10.40$\pm$0.019 &  9.39$\pm$0.002 &  9.14$\pm$0.003 &  8.73$\pm$0.003 &  8.04$\pm$0.002 &  5.21$\pm$0.010 &  0.05 &  -1.27 &  681 &  35. \\ 
21:04:20.57$\pm$0.07 &  50:16:08.26$\pm$0.13 &  13.83$\pm$0.025 &  12.96$\pm$0.006 &  12.35$\pm$0.004 &  11.14$\pm$0.005 &  10.73$\pm$0.007 &  10.08$\pm$0.009 &  9.20$\pm$0.008 &  6.37$\pm$0.027 &  1.41 &  -0.57 & \nodata & \nodata \\ 
21:04:28.81$\pm$0.39 &  50:18:24.20$\pm$0.07 &  13.48$\pm$0.002 &  12.70$\pm$0.001 &  12.25$\pm$0.001 &  11.61$\pm$0.004 &  11.38$\pm$0.004 &  10.98$\pm$0.008 &  10.14$\pm$0.008 &  7.44$\pm$0.036 &  0.20 &  -1.15 & \nodata & \nodata \\

\enddata
\tablenotetext{a}{Magnitudes are de-reddened when extinction value A$_{Ks}$ quoted.}
\tablenotetext{b}{A$_K$ determined from method outlined in $\S$5.}
\tablenotetext{c}{Slope $\alpha$ determined from IRAC photometry}
\tablenotetext{d}{Running number from \citet{herbig06}.}
\tablenotetext{e}{ EW[H$\alpha$] from \citet{herbig06}.}
\tablenotetext{f}{Slope $\alpha$ determined from [3.6], [5.8] and [24] photometry}
\tablenotetext{g}{Slope $\alpha$ determined from K-band to longest detected wavelength}
\end{deluxetable}

\clearpage


\begin{deluxetable}{llccccccccccccc}
\rotate
\tabletypesize{\scriptsize}
\setlength{\tabcolsep}{0.02in}
\tablecolumns{14}
\tablewidth{0pc}
\tablecaption{Objects with Mid Infrared Excesses (extragalactic contaminants removed): Transition Disks \label{classtdtable}}
\tablehead{
\colhead{RA$_{2000}$} & \colhead{DEC$_{2000}$} & \colhead{$J$}\tablenotemark{a} & \colhead{$H$}\tablenotemark{a} & \colhead{$K$}\tablenotemark{a} & \colhead{$[3.6]$}\tablenotemark{a} & \colhead{$[4.5]$}\tablenotemark{a} & \colhead{$[5.8]$}\tablenotemark{a} & \colhead{$[8.0]$}\tablenotemark{a} & \colhead{$[24]$}\tablenotemark{a} & \colhead{A$_{Ks}$}\tablenotemark{b}  & \colhead{$\alpha_{IRAC}$}\tablenotemark{c} & \colhead{IH$\alpha$}\tablenotemark{d} & \colhead{EW(H$\alpha$)}\tablenotemark{e} 
}

\startdata
21:03:27.22$\pm$0.76 &  50:13:00.70$\pm$0.41 &  11.63$\pm$0.001 &  10.97$\pm$0.001 &  10.72$\pm$0.024 &  10.65$\pm$0.003 &  10.52$\pm$0.003 &  10.35$\pm$0.008 &  9.80$\pm$0.006 &  6.60$\pm$0.022 &  0.88 &  -1.88 & \nodata & \nodata \\ 
21:03:52.07$\pm$0.18 &  50:14:18.65$\pm$0.36 &  11.19$\pm$0.001 &  10.59$\pm$0.017 &  10.55$\pm$0.021 &  10.62$\pm$0.005 &  10.71$\pm$0.009 &  10.62$\pm$0.030 &  10.49$\pm$0.056 &  4.83$\pm$0.027 &  0.59 &  -2.67 &  641 &  17. \\ 
21:03:57.80$\pm$0.27 &  50:12:31.18$\pm$0.21 &  14.00$\pm$0.003 &  13.37$\pm$0.001 &  13.17$\pm$0.002 &  12.94$\pm$0.007 &  12.83$\pm$0.008 &  12.82$\pm$0.046 &  12.53$\pm$0.083 &  7.92$\pm$0.107 &  0.38 &  -2.40 &  650 &  37. \\ 
21:04:02.41$\pm$0.44 &  50:07:39.27$\pm$0.58 &  6.58$\pm$0.011 &  5.95$\pm$0.015 &  5.75$\pm$0.007 &  5.64$\pm$0.002 &  5.86$\pm$0.002 &  5.52$\pm$0.002 &  5.45$\pm$0.005 &  3.84$\pm$0.018 &  0.85 &  -2.52 & \nodata & \nodata \\ 
21:04:08.69$\pm$0.03 &  50:14:25.80$\pm$0.20 &  12.79$\pm$0.001 &  12.19$\pm$0.001 &  12.07$\pm$0.001 &  11.79$\pm$0.014 &  11.64$\pm$0.010 &  11.38$\pm$0.073 &  10.56$\pm$0.048 &  8.17$\pm$0.225 &  0.36 &  -1.43 &  674 &  12. \\
21:04:12.64$\pm$0.01 &  50:16:00.98$\pm$0.24 &  14.56$\pm$0.007 &  13.59$\pm$0.002 &  12.80$\pm$0.002 &  11.46$\pm$0.006 &  11.50$\pm$0.008 &  10.54$\pm$0.008 &  10.19$\pm$0.014 &  7.83$\pm$0.068 &  0.48 &  -1.17 & \nodata & \nodata \\ 
21:04:23.84$\pm$0.46 &  50:13:05.78$\pm$0.40 &  14.06$\pm$0.002 &  13.35$\pm$0.001 &  13.03$\pm$0.002 &  12.53$\pm$0.007 &  12.49$\pm$0.007 &  12.30$\pm$0.021 &  11.98$\pm$0.048 &  9.29$\pm$0.129 &  0.03 &  -2.19 & \nodata & \nodata \\

\enddata
\tablenotetext{a}{Magnitudes are de-reddened when extinction value A$_{Ks}$ quoted.}
\tablenotetext{b}{A$_K$ determined from method outlined in $\S$5.}
\tablenotetext{c}{Slope $\alpha$ determined from IRAC photometry}
\tablenotetext{d}{Running number from \citet{herbig06}.}
\tablenotetext{e}{ EW[H$\alpha$] from \citet{herbig06}.}
\end{deluxetable}

\clearpage


\begin{deluxetable}{llccccccccccccc}
\tablecolumns{14}
\tablewidth{0pc}
\tabletypesize{\scriptsize}
\setlength{\tabcolsep}{0.02in}
\rotate
\tablecaption{Objects without Mid Infrared Excesses (extragalactic contaminants removed): Class III
\label{classIIItable}}
\tablehead{
\colhead{RA$_{2000}$} & \colhead{DEC$_{2000}$} & \colhead{$J$}\tablenotemark{a} & \colhead{$H$}\tablenotemark{a} & \colhead{$K$}\tablenotemark{a} & \colhead{$[3.6]$}\tablenotemark{a} & \colhead{$[4.5]$}\tablenotemark{a} & \colhead{$[5.8]$}\tablenotemark{a} & \colhead{$[8.0]$}\tablenotemark{a} & \colhead{$[24]$}\tablenotemark{a} & \colhead{A$_{Ks}$}\tablenotemark{b}  & \colhead{$\alpha_{IRAC}$}\tablenotemark{c} & \colhead{IH$\alpha$}\tablenotemark{d} & \colhead{EW(H$\alpha$)}\tablenotemark{e} 
}
\startdata
21:03:29.19$\pm$0.11 &  50:19:05.84$\pm$0.26 &  14.48$\pm$0.003 &  13.88$\pm$0.003 &  13.92$\pm$0.005 &  13.79$\pm$0.013 &  13.85$\pm$0.018 &  13.94$\pm$0.091 &  \nodata & \nodata &  0.29 & \nodata &  628 &  1.3 \\ 
21:03:32.17$\pm$0.68 &  50:10:55.25$\pm$0.63 &  14.52$\pm$0.002 &  13.92$\pm$0.001 &  13.66$\pm$0.002 &  13.19$\pm$0.008 &  13.15$\pm$0.010 &  13.10$\pm$0.042 &  13.05$\pm$0.073 & \nodata &  0.00 &  -2.68 &  629 &  3.9 \\ 
21:03:33.94$\pm$0.26 &  50:20:32.81$\pm$0.00 &  14.25$\pm$0.002 &  13.65$\pm$0.002 &  13.57$\pm$0.003 &  13.35$\pm$0.010 &  13.29$\pm$0.011 &  13.12$\pm$0.042 &  13.16$\pm$0.078 & \nodata &  0.12 &  -2.60 &  630 &  3.5 \\ 
21:03:46.60$\pm$0.27 &  50:10:43.50$\pm$0.25 &  12.36$\pm$0.001 &  11.69$\pm$0.001 &  11.43$\pm$0.001 &  11.19$\pm$0.005 &  11.19$\pm$0.004 &  11.12$\pm$0.014 &  11.14$\pm$0.037 & \nodata &  0.50 &  -2.77 &  636 &  11. \\ 
21:03:52.07$\pm$0.18 &  50:14:18.65$\pm$0.36 &  11.19$\pm$0.001 &  10.59$\pm$0.017 &  10.55$\pm$0.021 &  10.62$\pm$0.005 &  10.71$\pm$0.009 &  10.62$\pm$0.030 &  10.49$\pm$0.056 &  4.83$\pm$0.027 &  0.59 &  -2.67 &  641 &  17. \\ 
21:03:57.80$\pm$0.27 &  50:12:31.18$\pm$0.21 &  14.00$\pm$0.003 &  13.37$\pm$0.001 &  13.17$\pm$0.002 &  12.94$\pm$0.007 &  12.83$\pm$0.008 &  12.82$\pm$0.046 &  12.53$\pm$0.083 &  7.92$\pm$0.107 &  0.38 &  -2.40 &  650 &  37. \\ 
21:04:00.27$\pm$0.15 &  50:14:20.09$\pm$0.15 &  12.85$\pm$0.001 &  12.23$\pm$0.001 &  12.05$\pm$0.001 &  11.84$\pm$0.010 &  11.82$\pm$0.035 &  11.65$\pm$0.138 &  \nodata & \nodata &  0.29 & \nodata &  656 &  9.5 \\ 
21:04:02.57$\pm$0.06 &  50:14:08.70$\pm$0.40 &  12.45$\pm$0.001 &  11.85$\pm$0.001 &  11.73$\pm$0.001 &  11.63$\pm$0.015 &  11.60$\pm$0.026 &  11.36$\pm$0.063 &  11.22$\pm$0.156 & \nodata &  0.39 &  -2.33 &  662 &  4.5 \\ 
21:04:03.69$\pm$0.14 &  50:13:23.04$\pm$0.01 &  12.02$\pm$0.001 &  11.38$\pm$0.019 &  11.17$\pm$0.024 &  11.02$\pm$0.004 &  10.92$\pm$0.004 &  10.85$\pm$0.012 &  10.82$\pm$0.021 & \nodata &  0.25 &  -2.61 &  665 &  19. \\ 
21:04:04.16$\pm$0.09 &  50:15:01.24$\pm$0.20 &  11.37$\pm$0.001 &  10.77$\pm$0.017 &  10.67$\pm$0.021 &  10.55$\pm$0.003 &  10.57$\pm$0.004 &  10.46$\pm$0.016 &  10.53$\pm$0.048 & \nodata &  0.50 &  -2.80 &  667 &  1.8 \\ 
21:04:04.50$\pm$0.03 &  50:14:01.54$\pm$0.09 &  12.97$\pm$0.001 &  12.32$\pm$0.001 &  12.08$\pm$0.001 &  11.73$\pm$0.005 &  11.71$\pm$0.006 &  11.62$\pm$0.036 &  11.49$\pm$0.100 & \nodata &  0.21 &  -2.56 &  668 &  8.6 \\ 
21:04:05.58$\pm$0.01 &  50:15:58.35$\pm$0.10 &  12.04$\pm$0.001 &  11.44$\pm$0.060 &  11.29$\pm$0.024 &  11.07$\pm$0.005 &  11.15$\pm$0.006 &  11.01$\pm$0.013 &  11.05$\pm$0.034 & \nodata &  0.39 &  -2.78 &  669 &  8.6 \\ 
21:04:08.80$\pm$0.26 &  50:14:33.10$\pm$0.05 &  12.76$\pm$0.001 &  12.14$\pm$0.001 &  11.96$\pm$0.001 &  11.69$\pm$0.008 &  11.71$\pm$0.009 &  11.62$\pm$0.039 &  11.65$\pm$0.050 & \nodata &  0.22 &  -2.77 &  675 &  2.7 \\ 
21:04:14.19$\pm$0.05 &  50:09:52.96$\pm$0.10 &  13.28$\pm$0.001 &  12.68$\pm$0.001 &  12.60$\pm$0.001 &  12.37$\pm$0.006 &  12.38$\pm$0.006 &  12.31$\pm$0.022 &  12.10$\pm$0.037 & \nodata &  0.07 &  -2.53 &  677 &  5.1 \\ 
21:04:16.19$\pm$0.05 &  50:13:34.94$\pm$0.04 &  13.48$\pm$0.001 &  12.88$\pm$0.001 &  12.76$\pm$0.001 &  12.59$\pm$0.010 &  12.66$\pm$0.010 &  12.56$\pm$0.036 &  12.61$\pm$0.046 & \nodata &  0.27 &  -2.83 &  678 &  4.3 \\ 
21:04:18.92$\pm$0.09 &  50:20:28.47$\pm$0.39 &  12.33$\pm$0.002 &  11.73$\pm$0.001 &  11.69$\pm$0.001 &  11.37$\pm$0.004 &  11.61$\pm$0.006 &  11.28$\pm$0.017 &  11.12$\pm$0.023 & \nodata &  0.74 &  -2.44 &  679 &  35. \\ 
21:04:19.18$\pm$0.05 &  50:13:18.06$\pm$0.04 &  12.89$\pm$0.001 &  12.27$\pm$0.001 &  12.03$\pm$0.001 &  11.75$\pm$0.004 &  11.65$\pm$0.004 &  11.60$\pm$0.014 &  11.42$\pm$0.032 & \nodata &  0.00 &  -2.48 &  680 &  2.1 \\ 
21:04:22.07$\pm$0.08 &  50:15:44.40$\pm$0.20 &  11.68$\pm$0.001 &  11.08$\pm$0.017 &  10.95$\pm$0.022 &  10.85$\pm$0.004 &  11.00$\pm$0.004 &  10.88$\pm$0.011 &  10.75$\pm$0.013 & \nodata &  0.23 &  -2.69 &  682 &  2.1 \\ 
21:04:23.23$\pm$0.20 &  50:13:00.74$\pm$0.34 &  12.63$\pm$0.001 &  12.03$\pm$0.001 &  11.91$\pm$0.001 &  11.68$\pm$0.006 &  11.62$\pm$0.005 &  11.62$\pm$0.014 &  11.55$\pm$0.038 & \nodata &  0.26 &  -2.71 &  683 &  4.8 \\ 
21:04:25.63$\pm$0.47 &  50:13:40.91$\pm$0.04 &  13.33$\pm$0.001 &  12.66$\pm$0.001 &  12.42$\pm$0.001 &  12.25$\pm$0.006 &  12.22$\pm$0.009 &  12.06$\pm$0.039 &  12.27$\pm$0.065 & \nodata &  0.17 &  -2.83 &  684 &  8.1 \\

\enddata
\tablenotetext{a}{Magnitudes are de-reddened when extinction value A$_{Ks}$ quoted.}
\tablenotetext{b}{A$_K$ determined from method outlined in $\S$5.}
\tablenotetext{c}{Slope $\alpha$ determined from IRAC photometry}
\tablenotetext{d}{Running number from \citet{herbig06}.}
\tablenotetext{e}{ EW[H$\alpha$] from \citet{herbig06}.}
\end{deluxetable}

\clearpage

\begin{deluxetable}{rccc}
\tablecaption{Assumed Reddening Law from \citet{flaherty07}.\label{redtable}}
\tabletypesize{\scriptsize}
\tablewidth{0pc}
\tablehead{
\colhead{Waveband} &  \colhead{A$_{\lambda}$/A$_{Ks}$} 
}
\startdata
J & 2.5  \\
H & 1.55 \\
$K_{s}$ & 1.0 \\
Band 1 (3.6 $\mu$m) & 0.64 \\
Band 2 (4.5 $\mu$m) & 0.52 \\
Band 3 (5.8 $\mu$m) & 0.49 \\
Band 4 (8.0 $\mu$m) & 0.47 \\
\enddata
\end{deluxetable}

\clearpage

\begin{thebibliography}{lada06}
\bibitem[Allen et al.(2004)]{allen04} Allen, L. E., Calvet, N., D'Alessio, P., Merin, B., Hartmann, 
L., Megeath, S. T., Gutermuth, R. A., Muzerolle, J., Pipher, J. L., Myers, P. C.; Fazio, G. G. 2004, 
\apjs, 154, 363
\bibitem[Baraffe et al.(1998)]{baraffe98} Baraffe, I., et al., 1998, \aa, 337, 403
\bibitem[Clark(1986)]{clark86} Clark, F. O., 1986, A\&A, 164, L19
\bibitem[Carpenter(2001)]{carpenter01} Carpenter, J. M., 2001, \aj, 121, 2851
\bibitem[Dobashi et al.(1994)]{dobashi94} Dobashi, K., Bernard, J., Yonekura, 
Y.Fukui, Y., 2004, \apjs, 95, 419
\bibitem[D'Antona \& Mazzitelli(1997)]{DM97} D'Antona, F. \& Mazzitelli, I., 1997, Cool Stars 
in Clusters and Associations, ed. G. Micela \& R. Pallavicini (Firenzw: Soc. Astr. Italiana), 807
\bibitem[Elston et al.(2006)]{elston06} Elston, R., J., et al., 2006, \apj, 639, 816
\bibitem[Fazio et al.(2004)]{fazio04} Fazio, G., G., et al., 2004, \apj, 154, 39
\bibitem[Flaherty et al.(2007)]{flaherty07} Flaherty, K. et al., 2007, ApJ. in press
\bibitem[Glass(1999)]{glass99} Glass, I., Handbook of Infrared Astronomy (Cambridge: Cambridge Univ. Press)
\bibitem[Gutermuth et al.(2004)]{gutermuth04} Gutermuth, R. A., 
Megeath, S. T.,Muzerolle, J., Allen, L. E., Pipher, J. L., Myers, P. C., Fazio, 
G. G., 2004, \apjs, 154, 374
\bibitem[Gutermuth(2005)]{gutermuth05} Gutermuth, R. A., 2005, PhD, 
University of Rochester
\bibitem[Gutermuth et al.(2005)]{guter05} Gutermuth, R.A., Megeath, S.T., Pipher, J.L., Williams, 
J.P., Allen, L.E., Myers, P.C.\& Raines, S.N.., 2005, \apj, 632, 397
\bibitem[Gutermuth et al.(2007)]{guter07} Gutermuth, R.A. et al., 2007, in preparation.
\bibitem[Hartmann(1998)]{hartmann98} Hartmann, L., 1998, Accretion Processes in Star 
Formation, (Cambridge University Press: Cambridge), p. 124.
\bibitem[Herbig \& Bell(1988)]{herbig88} Herbig, G. H.\& Bell, K. R., 1988, Lick
Obs. Bull, n.1111, 1
\bibitem[Herbig \& Dahm(2006)]{herbig06} Herbig, G.H. \& Daum, S.E., 2006, \aj, 131, 1530.
\bibitem[Hernandez et al.(2007)]{hernandez07} Hernandez, J., Hartmann, L., Megeath, T., 
Gutermuth, R., Muzerolle, J., Calvet, N., Vivas, A.K., Briceno, C., Allen, L., Stauffer, J., Young, 
E., and Fazio, G., submitted to ApJ.
\bibitem[Hodapp(1994)]{hodapp94} Hodapp, K., 1994, \apjs, 94, 615
\bibitem[Indebetouw et al.(2005)]{indeb05} Indebetouw, R., Mathis, J. S., Babler, 
B.L., Meade, M. R., Watson, C., Whitney, B. A., Wolff, M. J., Wolfire, M. G., 
Cohen,M., Bania, T. M., and 10 coauthors, 2005, \apj, 619, 9311
\bibitem[Jenness et al.(1998)]{jlh98} Jenness, T., Lightfoot, J. F., Holland, W. 
S.1998, SPIE 3357, 548
\bibitem[Lada \& Lada(2003)]{ll03} Lada, C. J., \& Lada E. A., 2003, \araa, 41, 57
\bibitem[Lada et al.(2006)]{lada06} Lada, C.~J., et al., 2006, \aj, 131, 1574
\bibitem[Meyer et al.(1997)]{mch97} Meyer, M. R., Calvet, N., 
\& Hillenbrand, L. A., 1997, \aj, 114, 288
\bibitem[Muzerolle et al.(2004)]{muzerolle04} Muzerolle, J., Megeath, S. 
T.,Gutermuth, R. A., Allen, L. E., Pipher, J. L., Hartmann, L., Gordon, K. D., 
Padgett,D. L., Noriega-Crespo, A., Myers, P. C., Fazio, G. G., Rieke, G. H., 
Young, E. T.,Morrison, J. E., Hines, D. C., Su, K. Y. L., Engelbracht, C. W., 
Misselt, K. A.,2004, \apjs, 154, 379
\bibitem[Pipher et al.(2004)]{pipher04} Pipher, J. L., et al., 2004, SPIE 5487, 234 
\bibitem[Porras et al.(2003)]{porras03} Porras, A., Christopher, M., Allen, L., 
DiFrancesco, J., Megeath, S. T., Myers, P. C., 2003, \aj, 126, 
1916
\bibitem[Reach et al.(2005)]{reach05} Reach, W.T., Megeath, S.T., Cohen, M., Hora, J., Carey, S., 
Surace, J., Willner, S.P., Barmby, P., Wilson, G., Glaccum, W., Lowrance, P., Marengo, M., and 
Fazio, G.G., 2005, \pasp, 117, 978.
\bibitem[Ridge et al.(2003)]{ridge03} Ridge, N. A.. Wilson, T. L., Megeath, 
S. T.,Allen, L. E., Myers, P. C., 2003, \aj, 126, 286.
\bibitem[Siess et al.(2000)]{siess00} Siess, L., Dufour, E., \& Forestini, M. 2000, A\&A, 358, 593
\bibitem[MIPS Data Handbook(2005)]{mips_man} {\it Spitzer} Science Center. 2005, Multiband 
Imaging Photometer for {\it Spitzer} (MIPS) Data Handbook (Pasadena: SSC), 
http://ssc.spitzer.caltech.edu/mips/dh/
\bibitem[White \& Basri(2003)]{white03} White, R. J. \& Basri, G. 2003, \apj, 582, 1109
\bibitem[Whitney et al.(2003)]{whit03} Whitney, B. A., Wood, K., Bjorkman, J. 
E.,Cohen, M., 2003, \apj, 598, 1079
\bibitem[Whitney et al.(2004)]{whitney04} Whitney, B. A., et al., 2004, \apjs, 
154, 315
\bibitem[Wilson et al.(2003)]{wilson03} Wilson, J.  C., Eikenberry, S. 
S., Henderson,C.  P., Hayward, T.  L., Carson, J.  C.; Pirger, B. Barry, D.  J., Brandl, B.  R.,Houck, 
J.  R., Fitzgerald, G.  J., Stolberg, T. M., 2003, SPIE, 4841, 451
\end{thebibliography}
\end{document}